\documentclass[preprint,showpacs,showkeys,preprintnumbers,amsmath,amssymb,nofootinbib,aps]{revtex4-1}
\usepackage{amssymb}
\usepackage{amsmath}
\usepackage{graphicx}
\usepackage{dcolumn}
\usepackage{multirow}
\usepackage{bm}
\usepackage{epsfig}
\usepackage{xcolor}
\usepackage{slashed}
\usepackage[caption=false]{subfig}
\usepackage[all]{xy}
\usepackage[bookmarks=false]{hyperref}
\xyoption{import,color,arc}

\newcommand\beq{\begin{eqnarray}}
\newcommand\eeq{\end{eqnarray}}

\newcommand\nn{\nonumber}
 
\newcommand\Eq[1]{Eq.~\ref{eq:#1}}
\newcommand\Fig[1]{Fig.~\ref{fig:#1}}
\newcommand\Sec[1]{Sec.~\ref{sec:#1}}
\newcommand\Appendix[1]{Appendix~\ref{sec:#1}}

\newcommand\Tab[1]{Table~\ref{tab:#1}}

\newcommand\gcir{\gamma_{\bar{\chi}\chi,\,{\rm IR}}}
\newcommand\gpir{\gamma_{\bar{\psi}\psi,\,{\rm IR}}}
\newcommand\gir{\gamma_{\rm IR}}
\newcommand\Gpir{\gamma_{\bar{\psi}\psi,\,{\rm IR}}(2-\gamma_{\bar{\psi}\psi,\,{\rm IR}})}
\makeatletter

\newcommand{\Rmnum}[1]{\expandafter\@slowromancap\romannumeral #1@}
\makeatother

\begin{document}
\count\footins = 1000
\preprint{PNUPT-20/A01}

\title{Into the conformal window: multi-representation gauge theories
}

\begin{abstract}
We investigate the conformal window of four-dimensional gauge theories with fermionic matter fields in 
multiple representations. Of particularly relevant examples are the ultra-violet complete models with fermions in two distinct representations 
considered in the context of composite Higgs and top partial-compositeness. 
We first discuss various analytical approaches to unveil the lower edge of the conformal window 
and their extension to the multiple matter representations. 
In particular, we argue that the scheme-independent series expansion for the anomalous dimension of a fermion bilinear at an infrared fixed point, 
$\gcir$, 
combined with the conjectured critical condition, $\gcir = 1$ or equivalently $\gcir (2-\gcir)=1$, 
can be used to determine the boundary of conformal phase transition on fully physical grounds. 
In illustrative cases of $SU(2)$ and $SU(3)$ theories with $N_R$ Dirac fermions in various representations, 
we assess our results by comparing to other analytical or lattice results.
\end{abstract}

\author{Byung Su Kim}
\email{asviuw@gmail.com}

\author{Deog Ki Hong}
\email{dkhong@pusan.ac.kr}

\author{Jong-Wan Lee}
\email{jwlee823@pusan.ac.kr}
\affiliation{Department of Physics, Pusan National University, Busan 46241, Republic of Korea}





\date{\today}

\maketitle

\section{Introduction}
\label{sec:introduction}

The existence of a non-zero infrared (IR) fixed point in the renormalization-group (RG) beta function of 
asymptotically free gauge theories in four dimensions with a sufficient number of massless fermions $N_f$ for a given number of colors $N_c$ 
has been of particular interests recent years because of its potential application to phenomenological model buildings 
in the context of physics beyond the standard model (BSM), 
as well as its distinctive feature of conformal phase in contrast to the nonconformal phase as in Quantum Chromodynamics (QCD). 
A perturbative calculation  at the two-loop order in the weak coupling regime of such theories
finds an interacting IR fixed point \cite{Caswell:1974gg}, known as the Banks-Zaks (BZ) fixed point, 
named after their work on the phase structure of vector-like gauge theories with massless fermions at zero temperature \cite{Banks:1981nn}. 
As we vary the ratio of $N_f/N_c$, treated as a continuous variable, 
the IR fixed point either approaches zero,  at which the theory loses the asymptotic freedom and becomes trivial, 
or runs away into the strong coupling regime where the perturbative expansion breaks down.  For sufficiently small values of the ratio we expect the theory is in a chirally broken phase, 
that implies the presence of a zero-temperature quantum phase transition between the conformal and chirally broken phases at a critical value of the ratio. 
A finite range of the number of flavors for which the theory has a non-zero IR fixed point is called {\it conformal window}, 
and the chirally-broken theories near the phase transition are expected to have quite different IR dynamics, compared to QCD-like theories. 

Near-conformal dynamics is ubiquitous in BSM models of which 
the underlying ultraviolet (UV) theory is a novel strongly coupled gauge theory. 
One of its crucial features is a large anomalous dimension of relevant composite operators. 
In walking technicolor 
which is supposed to have a slowly evolving coupling and thus provides a large separation in the scale between the chiral symmetry breaking $\Lambda_{\chi}$ 
and the confinement $\Lambda_{\rm TC}$, 
a large anomalous dimension of the chiral order-parameter
is expected to achieve the dynamical electroweak symmetry breaking while naturally avoiding constraints from the flavor physics 
\cite{Holdom:1981rm,Yamawaki:1985zg,Akiba:1985rr,Appelquist:1986an}. 
Similarly in the composite Higgs models, which realize both the pseudo Nambu-Goldstone bosons (pNGB) Higgs \cite{Kaplan:1983fs,Kaplan:1983sm,Dugan:1984hq}
and the partial compositeness for the top quark \cite{Kaplan:1991dc}, a large anomalous dimension of baryonic operators linearly coupled to the standard model (SM) top quark 
is assumed to explain the relatively large mass of top quark, compared to other quarks. 
This idea was originally proposed in the framework of warped extra dimensions \cite{Contino:2003ve,Agashe:2004rs}, 
and the corresponding minimal models have been extensively studied in various phenomenological aspects at the level of effective theories (see \cite{Contino:2010rs,Panico:2015jxa} for reviews, and references therein). 
However, it is relatively recent to consider the realistic candidates for the four-dimensional UV complete models based on strongly coupled gauge theories, containing two different representations of fermionic matter fields 
\cite{Barnard:2013zea,Ferretti:2013kya,Ferretti:2016upr}. 
\footnote{
Even if they are not near conformal, 
the two-representation composite Higgs models usually have additional light and non-anomalous pseduo-scalars that have interesting phenomenological signatures
at the colliders, as studied in~\cite{Cacciapaglia:2019bqz}.
}
The anomalous dimension of the baryonic operators for the top-partner
was calculated 
at one-loop in the perturbative expansion 
for some of these models \cite{DeGrand:2015yna} 
and also for the relevant IR-conformal theories \cite{BuarqueFranzosi:2019eee}. 
Furthermore, substantial efforts have been devoted to investigate the low-energy dynamics of 
this kind of theories from the first-principle Monte Carlo (MC) lattice calculations, 
in particular for 
$SU(4)$ \cite{Ayyar:2017qdf,Ayyar:2018zuk,Ayyar:2018ppa,Ayyar:2018glg,Ayyar:2019exp,Cossu:2019hse} 
and $Sp(4)$ \cite{Bennett:2017kga,Lee:2018ztv,Bennett:2019jzz,Bennett:2019cxd} gauge theories. 

The other common non-trivial features of near-conformal gauge theories are the emergence of a light scalar resonance. 
Such a new degree of freedom at low energy may be identified as a dilaton arising from the spontaneous breaking of scale symmetry, 
and can be used to extend the Higgs sector of the standard model of particle physics 
\cite{Hong:2004td,Dietrich:2005jn,Hashimoto:2010nw,Appelquist:2010gy,Vecchi:2010gj,
Chacko:2012sy,Bellazzini:2012vz,Abe:2012eu,Eichten:2012qb,Hernandez-Leon:2017kea,Hong:2017smd}. 
Interestingly, recent lattice studies of $SU(3)$ gauge theories with $8$ fundamental Dirac fermions \cite{Aoki:2014oha,Appelquist:2016viq,Aoki:2016wnc,Appelquist:2018yqe}, 
as well as $2$ two-index symmetric Dirac fermions 
\cite{Fodor:2012ty,Fodor:2015vwa,Fodor:2016pls,Fodor:2017nlp,Fodor:2019vmw}, 
performed with moderate sizes of the fermion mass 
found a relatively light scalar in the spectrum. 
There have been several attempts to analyze these results within a low-energy effective field theory (EFT) 
\cite{Golterman:2016lsd,Golterman:2016cdd,Golterman:2018mfm,Appelquist:2017wcg,Appelquist:2017vyy,Appelquist:2019lgk}. 
The dilaton potential also inherently possesses the possibility of a strong first-order phase transition at a finite temperature, 
needed for the electroweak baryogenesis \cite{Bruggisser:2018mus,Bruggisser:2018mrt} 
and the supercooled universe \cite{Konstandin:2011dr,Iso:2017uuu,vonHarling:2017yew,Baratella:2018pxi} in the context of  composite Higgs scenarios. 

While phenomenological model buildings could be carried out under some working assumptions, utilizing the qualitative features of near-conformal dynamics at low energy, 
in order to explore its properties fully it is necessary to perform quantitative studies from the underlying strongly-coupled gauge theories. 
As mentioned above, lattice MC calculations are highly desired in this respect, 
where most of the modern technologies developed for the lattice QCD can be applied without additional difficulties. 
However, lattice calculations are expensive and thus practically not suitable to explore all the possibilities in the theory space 
at arbitrary numbers of $N_c$ and $N_f$. 
Therefore, any analytical calculations that map out the conformal window are greatly welcome to find the most promising UV models of the near-conformal dynamics. 
While various analytical proposals are made in the literature \cite{Appelquist:1999hr,Ryttov:2007cx,Pica:2010mt} 
besides the traditional Schwinger-Dyson analysis, 
we propose in this paper to use the critical condition on the anomalous dimension of 
a fermion bilinear operator at an IR fixed point, $\gamma_{\rm IR}=1$ or equivalently $\gamma_{\rm IR} (2-\gamma_{\rm IR})=1$, for the conformal phase transition to occur. 
We do not claim the originality of this idea: 
in Ref. \cite{Appelquist:1998rb} the conformal window of $SU(N)$ gauge theory with $N_f$ fundamental fermions 
was described by using the critical condition, calculating the anomalous dimension in the loop expansion. 
We instead emphasize that it becomes an alternative method to map out the conformal window 
in a scheme-independent way if we adopt the series expansion of $\gamma_{\rm IR}$ recently developed by 
Ryttov and Shrock \cite{Ryttov:2016hdp,Ryttov:2016asb,Ryttov:2016hal,Ryttov:2017toz,Ryttov:2017kmx,Ryttov:2017dhd,Ryttov:2018uue}. 
We find that this method is particularly useful to discuss the sequential condensates of fermions in different representations, which are expected in the near-conformal theories~\footnote{The chiral symmetry breaking of fermions in one representation might induce the chiral symmetry breaking of other representation through the gauge interactions. For the near conformal dynamics, however, because the gauge coupling remains almost constant for a wide range of scales between the chiral symmetry breaking and the dynamical mass generation, such effect is negligible. Only after the chirally-broken fermions decouple, the gauge coupling becomes strong enough  to break the other chiral symmetry.}.
Although we restrict our attention to the case with fermions in the two different representations, relevant to the composite Higgs models 
as summarized in Ref. \cite{Belyaev:2016ftv}, 
the methodology discussed in this work can be straightforwardly extended to the case with fermions in any number of representations. 

The paper is organized as follows. 
In \Sec{CW} we provide some general remarks on the conformal window for a generic nonabelian gauge theory with fermions in multiple representations. 
We then describe several analytical methods, studied in the literature to determine the lower bound of the conformal window.  We also revisit the critical condition of the anomalous dimension of the fermion bilinear operators 
for chiral symmetry breaking. 
In \Sec{SIPT} we briefly review the scheme-independent calculation of $\gir$ 
for gauge theories with fermions in one or two different representations, 
and determine the lower bound of the conformal window in the exemplified cases of $SU(2)$ and $SU(3)$ gauge theories 
with $N_R$ Dirac fermions in various representations. 
We assess our results by comparing to several scheme-dependent calculations 
as well as other analytical or lattice results. 
We also discuss the convergence of the scheme-independent expansion for the critical condition. 
In \Sec{CHM} we present our main results on the conformal window for the two-representation gauge theories, 
relevant to the composite Higgs models and the top partial-compositeness. 
We present some results on the group invariants used to compute the coefficients of the scheme-independent series expansions in \Appendix{group_inv}, 
and the lower-order results for the conformal window in \Appendix{loops}. 
Finally, we conclude by summarizing our findings in \Sec{conclusion}. 

\section{Conformal window: analytical approaches}
\label{sec:CW}

We start by providing a general remark on the conformal window of four-dimensional gauge theories 
containing fermionic matter in the multiple representations with $\{N_{R_i}\}$, $i=1,\,2,\,\cdots,\, k$, 
denoting a set of the number of flavors in the representation $R_i$. 
In a small coupling regime, the perturbative beta function is given in powers of the gauge coupling $\alpha=g^2/(4\pi)$ as 
\beq
\beta(\alpha) \equiv \frac{\partial \alpha(\mu)}{\partial \ln \mu}=-2\alpha \sum_{\ell=1}^\infty b_\ell \left(\frac{\alpha}{4\pi}\right)^\ell,
\eeq
where $b_\ell$ is the $\ell$-loop coefficient and $\mu$ is the renormalization scale. 
The coefficients of the lowest two terms, $b_1$ \cite{Gross:1973id,Politzer:1973fx} and $b_2$ \cite{Caswell:1974gg}, are renormalization 
scheme-independent and given as
\beq
b_1=\frac{11}{3}C_2(G)-\frac{4}{3}\sum_{i=1}^k N_{R_i} T(R_i),
\label{eq:b1}
\eeq
and
\beq
b_2=\frac{34}{3}C_2(G)^2-\frac{4}{3}\sum_{i=1}^k \left(5 C_2(G)+3 C_2(R_i)\right) N_{R_i} T(R_i).
\label{eq:b2}
\eeq
The generators in the representation $R_i$ of an arbitrary gauge group $G$ are denoted by 
$T_{R_i}^a$, $a=1,\cdots,d(G)$, where $d(G)$ is the dimension of the adjoint representation. 
The trace normalization factor $T(R_i)$ and the quadratic Casimir $C_2(R_i)$ are defined through 
${\rm Tr}[T_{R_i}^a T_{R_i}^b]=T(R_i)\delta^{ab}$ and $T_{R_i}^a T_{R_i}^a=C_2(R_i) I$, respectively. 
These two group-theoretical factors are related by $C_2(R_i) d(R_i) = T(R_i) d(G)$. 
Note that $b_\ell$ with $\ell \geq 3$ are known to be scheme-dependent. 
For the general discussions in this and the following section 
we use $N_{R_i}$ for the number of Dirac flavors in the representation $R_i$. 

As far as the UV completion is concerned, we require the theory is asymptotically free 
or $b_1>0$. 
(We do not consider the scenarios of UV safety in this work.) 
This condition leads to the maximum number of flavors above which we lose the asymptotic freedom. 
For a single representation, it is given by $N_R=\left[N_R^{\rm AF}\right]$, a largest integer but smaller than $N_R^{\rm AF}$ with 
\beq
N_R^{\rm AF}=\frac{11 C_2(G)}{4 T(R)},
\label{eq:AF}
\eeq
while for the $k$ number of representations they span the points on the ($k-1$)-dimensional surface in the space of $\{ N_{R_i}\}$ with $i=1,\,2,\,\cdots,\,k$, 
that satisfies $b_1=0$.
Since most of the discussion below is independent of whether the representations are multiple or not, 
we simply consider a single representation $R$ unless multiple representations are explicitly needed. 
For a sufficiently small and positive value of $b_1$, 
the theory develops a non-zero IR fixed point (BZ fixed point), if the two-loop coefficient $b_2$ is negative, 
\beq
\alpha_{\rm BZ} \simeq -4\pi \frac{b_1}{b_2}. 
\label{eq:alpha_BZ}
\eeq
This perturbative analysis suggests the existence of the conformal theory at small coupling $\alpha=\alpha_{\rm BZ}$ for  $N_R$ sufficiently large but still smaller than $N_{R}^{\rm AF}$ so that $b_1 \ll 1$. 

As we decrease $N_R$, however, $\alpha_{\rm BZ}$ increases in general and at some point the two-loop result is no longer reliable. 
Higher order corrections should be then included to extend the perturbative two-loop results, 
but are largely limited due to its scheme dependence. 
If $\alpha_{\rm BZ} \gtrsim \mathcal{O}(1)$, the perturbative expansion will break down. Furthermore
one has to take into account the nonperturbative effects of the IR dynamics.
Nevertheless, if we keep decreasing $N_R$, the (negative) slope of the beta function at UV becomes large enough 
so that the theory becomes strongly coupled at low energy and eventually falls into the chirally broken phase. 
One of the extreme case is pure Yang-Mills at $N_R=0$, which is confining therefore nonconformal. 
We therefore expect that there is a finite range of the number of flavors $N_R$, 
namely a conformal window (CW), 
where the theory is conformal in IR. 
While the upper bound of CW is identical to that for losing the asymptotic freedom, $\left[N_{R}^{\rm AF}\right]$,
its lower bound $N_{R}^c$ is not easy to determine because of the difficulties mentioned above. 
In the following sections, we briefly discuss several analytical but approximate approaches being used to determine the lower bound $N_{R}^c$ of CW. 
Of our particular interest is the one obtained from the critical condition of the anomalous dimension of fermion bilinear operators, discussed 
in \Sec{CC}. 

\subsection{$2$-loop beta function}
\label{sec:two_loop}
A naive estimation of the lower bound for CW comes from the criterion that the coupling at the BZ fixed point $\alpha_{\rm BZ}$ blows up to infinity. 
If we neglect the scheme-dependent higher order corrections, from the $2$-loop beta function we find the condition $b_2=0$ for the lower bound. 
Analogous to the upper bound of CW the solution lives on the $(k-1)$-dimensional surface for the multiple representations $\{R_1,\,R_2,\,\cdots ,\, R_k\}$. 
For a single representation, we obtain a simple expression
\beq
N_R^{c, {\rm 2-loop}}=\frac{17 C_2(G)^2}{T(R)\left[10 C_2(G)+6 C_2(R)\right]}.
\label{eq:nc_twoloop}
\eeq

\subsection{(traditional) Schwinger-Dyson approach with the ladder approximation}
\label{sec:SD}

It is well known that the Schwinger-Dyson (SD) gap equation for the fermion propagator in the ladder (or rainbow) approximation 
yields the critical coupling, a minimal coupling strength required to trigger the chiral symmetry breaking, given as
\beq
\alpha_{c} = \frac{\pi}{3C_2(R)}. 
\label{eq:critical_alpha}
\eeq
The traditional way to determine the number of flavors for the onset of chiral symmetry breaking 
is to equate the $2$-loop IR fixed point, $\alpha_{\rm BZ}$ in \Eq{alpha_BZ},  with $\alpha_c$, 
which gives for the single representation $R$ 
\beq
N_R^{c,{\rm SD}}=\frac{C_2(G)(17 C_2(G)+66 C_2(R))}{T(R) (10 C_2(G)+30 C_2(R))}.
\label{eq:nc_2_loop}
\eeq
Note that the critical coupling is inversely proportional to $C_2(R)$. 
Furthermore, in near conformal theories with fermions in the multiple representations, 
one expects the fermions form chiral condensates sequentially, if they do: 
the fermions in the representation having the largest value of $C_2(R)$, denoted by $R_1$, 
would first be integrated out from the theory at some scale $\Lambda_1$ when they develop a dynamical mass. 
For $\mu < \Lambda_1$ the beta function will change to include only the low-energy effective degrees of freedom except the fermions in $R_1$, 
and this procedure will sequentially occur as we decrease the scale $\mu$~\cite{Ryttov:2010hs}.\footnote{
The sequential IR evolution of fermion condensates should be understood 
as a conjecture, 
since no rigorous proof such as lattice simulations for this kind of theories are performed yet. 
Recently  the $SU(4)$ lattice gauge theory 
with $2$ fundamental and $2$ two-index antisymmetric Dirac fermions is studied at finite temperature~\cite{Ayyar:2018glg} to find that 
chiral symmetry breaking and color confinement occur at the same critical temperature  for the fermions considered. 
As we will see in \Sec{CHM}, however, this theory is expected to be located deep inside the chirally broken phase, far away from the conformal window.
} 
In this case, therefore, the theory will leave the conformal window when 
the IR coupling $\alpha_{\rm IR}$ exceeds the critical coupling $\alpha_{R_1,\, c}$ for $R_1$. 

\subsection{All-orders beta function}
\label{sec:BF}

The coefficients of the lowest two terms in the perturbative beta function do not depend on the renormalization scheme, 
so does the lower bound, $N_R^c$, discussed in the previous two sections. 
While this is no longer true if one considers higher order terms in the beta function, 
it is believed that there exists a certain scheme such that all higher order terms ($\ell \geq 3$) vanish or 
at least the beta function is written in a closed form. 
Along the line of this idea all-orders beta functions are suggested in Refs. \cite{Ryttov:2007cx,Pica:2010mt}, 
inspired by the Novikov-Shifman-Vainshtein-Zakharov (NSVZ) beta function for supersymmetric theories \cite{Novikov:1983uc}. 
The conjectured beta function for generic gauge theories with Dirac fermions in multiple representations, 
proposed in \cite{Ryttov:2007cx}, is written in the following form
\beq
\beta^{\rm all-orders}(\alpha)=-\frac{\alpha^2}{2\pi}
\frac{b_1-\frac{2}{3}\sum_{i=1}^k T(R_i) N_{R_i} \gamma_{R_i} (\alpha)}
{1-\frac{\alpha}{2\pi}C_2(G)\left(1+\frac{2 b'_1}{b_1}\right)}
\eeq
where $\gamma_{R_i}$ is the anomalous dimension of a fermion bilinear for a given representation $R_i$, 
$b'_1=C_2(G)-\sum_{i=1}^k T(R_i) N_{R_i}$, and $b_1$ is defined in \Eq{b1}. 
For a single represenation $R$, using the leading-order expression for $\gamma_R(\alpha)$, 
this beta function reproduces the (universal) perturbative two-loop results. 
Note that the IR fixed point is determined by taking $\beta^{\rm all-orders}(\alpha)=0$, 
which is physical in the sense that it only involves scheme-independent quantities 
such as the anomalous dimension $\gamma_{R_i}$. 

In the case of the single representation $R$, the anomalous dimension at the IR fixed point is given by
\beq
\gamma_{\rm IR}=\frac{11 C_2(G)}{2 T(R) N_R} -2.
\label{eq:BF_gIR_single}
\eeq
The lower bound of CW is typically determined by taking $\gamma_{\rm IR}=2$, implied from the unitarity~\cite{Mack:1975je}. 
Unfortunately, $\gamma_{\rm IR}$ determined by \Eq{BF_gIR_single} turns out to be inconsistent with the perturbative result at the IR fixed point. 
A modified version of the all-orders beta functions that resolves the inconsistency was later proposed  in \cite{Pica:2010mt}. 
But now the unitarity condition leads to too small values of $N_R^c$ for the lower edge of the conformal window, 
e.g. smaller than the value obtained from \Eq{nc_twoloop}, 
which shows the unitarity condition is too weak. 

In contrast to the case of a single representation, 
the all-orders beta function provides no simple expressions for 
the anomalous dimensions at the IR fixed point as in \Eq{BF_gIR_single}: 
we rather have
\beq
\frac{2}{11}\sum_{i=1}^k T(R_i) N_{R_i} (2+\gamma_{R_i,\,{\rm IR}}) = C_2(G).
\eeq
As for the single representation, the lower bound may be obtained by applying the unitarity condition 
to all the representations, $\gamma_{R_i,\,{\rm IR}}=2$ with $i=1,\,2,\, \cdots ,\,k$. 
However, this approach does not give any informations on the aforementioned sequencial chiral symmetry breaking 
near the lower edge of the conformal window. 
Note that in general the anomalous dimensions of the fermion bilinears in different representations 
are expected to have different values at the IR fixed point. 

\subsection{Critical condition for the anomalous dimension of a fermion bilinear}
\label{sec:CC}

The critical coupling in \Eq{critical_alpha} being equal to $\alpha_{\rm BZ}$ has been widely used to estimate the phase boundary 
of the conformal window. 
However, the essence of the critical condition is actually hidden in the anomalous dimension of the fermion bilinear at the IR fixed point 
$\gir$ \cite{Appelquist:1988yc,Cohen:1988sq,Appelquist:1998rb}. 
To see this, let us recall the Schwinger-Dyson equation for the massless fermions, 
where the full inverse propagator in the momentum space is given as
\beq
iS^{-1}(p)=Z(p)\slashed{p}-\Sigma(p), 
\eeq
with $Z(p)$ and $\Sigma(p)$ being the wave-function renormalization constant and the self-energy function, respectively. 
In the Euclidean space the SD equation in the ladder approximation leads to 
the integral gap equation
\beq
\Sigma(p)=3 C_2(R) \int \frac{d^4k}{(2\pi)^4} 
\frac{\alpha((k-p)^2)}{(k-p)^2} \frac{\Sigma(k^2)}{Z(k^2) k^2 + \Sigma^2(k^2)}.
\label{eq:gap}
\eeq
In the Landau gauge, $Z(k^2)=1$, 
this equation can be linearized by neglecting $\Sigma^2(k^2)$ 
in the regime of sufficiently large momenta. 
The slowly varying coupling $\alpha(\mu)\approx \alpha_{\rm IR}$, which is the key assumption of near-conformal dynamics, 
further simplifies \Eq{gap} and one obtains two scale-invariant solutions for $\Sigma(p^2)$ of the form, 
$(p^2)^{-\gir/2}$, in the deep UV 
with \cite{Appelquist:1988yc} 
\beq
\gir(2-\gir)=\frac{\alpha_{\rm IR}}{\alpha_c},
\label{eq:critical_gamma}
\eeq
where $\alpha_c$ is given in \Eq{critical_alpha}. 
For $\alpha_{\rm IR} < \alpha_c$ the two solutions 
can be understood as the RG running of a renormalized mass $m(\mu)$ and a fermion bilinear operator $\bar{\chi}\chi(\mu)$ 
within the operator product expansion (OPE) at large Euclidean momentum. 
In this case no solution is found for non-vanishing chiral condensate with a vanishing mass term, 
indicating that no spontaneous chiral symmetry breaking occurs. \cite{Cohen:1988sq} 

For $\alpha_{\rm IR} \geq \alpha_c$ both solutions show the same $p$ dependence up to a phase difference 
(at $\alpha_{\rm IR}=\alpha_c$, $\Sigma(p) \sim (1/p^2)^{-1/2}$), 
and the OPE identification becomes obscure. 
As discussed in details in Ref.~\cite{Cohen:1988sq}, 
in fact, this situation can be described by a underdamped anharmonic oscillator that corresponds to 
spontaneous symmetry breaking. 
In the same paper, the authors showed that the generic feature of the transition between conformal and chirally broken phases 
imposed by the critical condition, $\alpha_{\rm IR} = \alpha_c$ or equivalently $\gir=1$, persists beyond the ladder approximation, 
though the details such as the value of $\alpha_c$ may change. 
Utilizing the critical condition on the anomalous dimension instead of the gauge coupling makes more sense to determine the phase boundary 
between conformal and non-conformal phases since it is physical and thus free from the renormalization scheme-dependency.

Interestingly, the critical condition derived from the truncated SD analysis is in agreement with the conjectured mechanism 
responsible for the zero-temperature conformal phase transition, 
featured by an annihilation of IR and UV fixed points \cite{Kaplan:2009kr}. 
As we approach the lower edge of the conformal window from above, in particular, 
the mass dimension of the operator $\bar{\chi}\chi$ at IR fixed point $\Delta_+$ decreases, while that of the counterpart at UV fixed point $\Delta_-$
increases, and becomes identical to each other at the transition point to give $\Delta_+=\Delta_- = 2$ in the four-dimensional spacetime. 
In a simplified holographic model \cite{Klebanov:1999tb} the loss of conformality occurs 
when the mass squared of a bulk scalar in a higher dimensional theory 
violates the Breitenlohner-Freedman (BF) bound, 
and the AdS/CFT correspondence implies that the dimension of the fermion bilinear operator 
is equal to $2$ at the conformal phase transition. 
As we discussed above if we cross the phase boundary from inside of the conformal window, 
the truncated SD equations no longer have the valid scale-invariant solutions. 
Analogously, the solutions to the beta function describing the fixed point merger become complex 
and give arise to a mass gap $m \sim \Lambda_{\rm UV}\, \textrm{exp}\left(-c/\sqrt{\alpha_{\rm IR}-\alpha_c}\right)$ with $c>0$ \cite{Kaplan:2009kr}. 
Recently, it is argued that such IR dynamics (walking dynamics), slightly below the conformal window, 
could be analyzed by using 
conformal perturbation theory in the vicinity of a complex pair of fixed points \cite{Gorbenko:2018ncu}. 

At the onset of chiral symmetry breaking, $\alpha_{\rm IR}=\alpha_c$, 
the solution to \Eq{critical_gamma} is equivalent to the condition $\gamma_{\rm IR}=1$, 
which results in the nonperturbative, gauge invariant and scheme-independent definition for the critical condition. 
In this work, we attempt to calculate $\gamma_{\rm IR}$ in a perturbative but scheme-independent manner. 
Note that, although both conditions of $\gir(2-\gir)=1$ in \cite{Appelquist:1988yc} and $\gir=1$ are not distinguishable in full theory, 
they provide two different definitions when the perturbative expansion is truncated at a finite order. 
Practically the former condition has been adopted since the leading-order expression of $\gir$ 
leads to the critical coupling $\alpha_c$ in \Eq{critical_alpha}, 
and the higher order estimates in the modified minimal subtraction scheme ($\overline{{\rm MS}}$) were studied in Ref. \cite{Appelquist:1998rb}. 
In general, this perturbative approach suffers from the scheme-dependency when higher-order terms ($\ell \geq 3$) are concerned. 
As we will discuss in \Sec{SIPT}, however, it turns out that we are able to circumvent this problem 
by adopting the scheme-independent series expansions for $\gir$ at the IR fixed-point. 
We will also discuss the convergence of the perturbative expansions for both definitions of the critical condition. 

We have so far restricted our attention to gauge theories with fermions in a single representation. 
In order to account for the two-representation theories relevant to composite Higgs models with partial compositeness, 
we should extend the critical condition discussed above to the case of fermions in multiple representations. 
Analogous to the critical couplings considered in the traditional SD approach in \Sec{SD}, 
the anomalous dimensions of fermions in the different representations give rise to different values at the IR fixed point \cite{Ryttov:2010hs}. 
The fermion representation, say $R_1$, whose anomalous dimension reaches unity first,  develops a non-vanishing fermion condensate first 
and thus provides the critical condition for the whole theory, unless the effective theory after integrating out the fermions in the representation $R_1$ 
does have an IR fixed-point. 
Keeping an eye on the sequence of critical conditions in the case of multiple representations, 
we impose the following conditions to determine the lower edge of conformal window, 
\beq
\textrm{Max}\left[\{\gamma_{R_i,\,{\rm IR}}\}\right] \equiv 1, ~~~\textrm{or}~~
\textrm{Max}\left[\{\gamma_{R_i,\,{\rm IR}}(2-\gamma_{R_i,{\rm IR}})\}\right] \equiv 1.
\label{eq:cc_multi_irreps}
\eeq
Again, we note that these two conditions are equivalent if all orders  are considered in the perturbative expansion, 
but they could in general result in two different sets of $\{N_{R_i}^c\}$ if the expansion is truncated at a finite order in the perturbative expansion. 

\subsection{Comparison between various analytical approaches}
\label{sec:method_comparison}
We conclude this section by comparing the analytical approaches to determine the lower edge of the conformal window. 
For convenience let us use the abbreviations $2$-loop, SD, BF and $\gamma$CC to denote 
the methods discussed in Sections \ref{sec:two_loop}, \ref{sec:SD}, \ref{sec:BF} and \ref{sec:CC}, respectively. 
Both the $2$-loop and SD methods use the gauge coupling at the BZ fixed-point, taken to be 
$\alpha_{\rm BZ} = \infty$ and $\alpha_{\rm BZ}=\alpha_c$, respectively.
One could extend these methods to higher-loops, but one then immediately encounters the complication of scheme dependence. 
On the other hands, the BF and $\gamma$CC methods rely on the anomalous dimension of a fermion bilinear at an IR fixed point $\gir$ to determine the conformal window on physical grounds. 
If we restrict ourselves to the case of a single representation, BF provides the exact value of $\gir$ at the IR fixed point in a scheme-independent way. 
For the onset of chiral symmetry breaking one typically chooses $\gir=2$ inspired by supersymmetric theories. 
To use $\gamma$CC one needs the value of $\gir$, which could be obtained perturbatively. 
As we will discuss in details in the following sections, one can still maintain the scheme independence of $\gamma$CC beyond the $2$-loop orders 
by incorporating the scheme-independent series expansions, proposed in Ref.~\cite{Ryttov:2016hdp}. 

We now turn our attention to the multiple representations. 
The $2$-loop method can be easily extended to the case of multiple representations by taking $b_2$ in \Eq{b2} to be zero. 
In contrast to the case of a single representation, BF provides neither the values of $\gamma_{R_i,\,{\rm IR}}$ nor 
the sequence of chiral symmetry breaking near the conformal window. 
However, one might still estimate the lower bound of CW by taking $\gamma_{R_i,\,{\rm IR}}=2$ for all the representations. 
In the cases of SD and $\gamma$CC one can use the dynamical results of $\alpha_{R_i,\,c}$ and $\gamma_{R_i,\,{\rm IR}}$ 
as they have different values for different representations. 
Assuming the theory falls into the chirally broken phase away from the conformal window, 
the representation having the maximum values of $\gamma_{R_i,\,{\rm IR}}$ and $\alpha_{R_i,\,c}^{-1}$ 
determines the lower edge of the conformal window. We note here that all of the above discussions are limited in principle since the nonperturbative effects of the strong dynamics in IR are not considered.

\section{Scheme-independent determination of conformal window using $\gamma$CC }
\label{sec:SIPT}

In this section we briefly review the scheme-independent (SI) series expansion of physical quantities 
at an IR fixed-point in asymptotically free gauge theories with fermions in a single representation, and its extension to multiple representations. 
We present only the essential ingredients, focusing mainly on the calculation of the anomalous dimensions of fermion bilinear operators, 
needed for our discussions. 
The great details, including how to calculate other physical quantities, can be found in a series of works done 
in~\cite{Ryttov:2016hdp,Ryttov:2016asb,Ryttov:2016hal,Ryttov:2017toz,Ryttov:2017kmx,Ryttov:2017dhd,Ryttov:2018uue,Gracey:2018oym}. 
We then describe how to determine the conformal window from the critical condition for the anomalous dimension $\gamma$CC, using this new technique. 
In the illustrative examples of $SU(2)$ and $SU(3)$ gauge theories with $N_R$ Dirac fermions in various representations, 
we discuss the consequence of the critical condition, written in two different forms in \Eq{cc_multi_irreps}, truncated at a finite order in the SI expansion, 
and compare our results to the various scheme-dependent expansions and other analytical (but approximate) approaches 
together with non-perturbative lattice results. 

\subsection{Scheme independent series expansion of $\gamma_{\rm IR}$}
\label{sec:SI}

A series expansion of the anomalous dimension of a fermion bilinear, made of 
fermions in the representation $R$, at an IR fixed-point, in  terms of the scheme-independent variable $\Delta_{R} \equiv(N_{R}^{\rm AF}-N_{R})$ 
has been proposed by Ryttov \cite{Ryttov:2016hdp} to write 
\beq
\label{eq:gamma_SI}
\gamma_{\rm IR}(\Delta_{R})=\sum_{i=1}^{\infty} c_{i}(\Delta_{R})^{i}.
\eeq
The coefficients of each term are clearly scheme-independent because the anomalous dimension in the left-handed side is physical, 
scheme-independent, and $N_R^{\rm AF}$ is defined from the scheme-independent one-loop beta function as in \Eq{AF}. 
Furthermore it has been shown that the $i$-th order coefficient $c_{i}$ depends only on the coefficients of the beta function 
and the anomalous dimension at the $(i+1)$-th and $i$-th loops, evaluated at $\Delta_{R}=0$, respectively. 
Namely, there are no higher-loop corrections to the coefficient $c_{i}$, 
though the coefficient is scheme-independent. 

To determine the coefficients $c_i$ in \Eq{gamma_SI} 
we first note that the coupling at the IR fixed-point may be expanded as 
\begin{equation}
\frac{\alpha_{\rm IR}}{4\pi}=\sum_{j=1}^{\infty}a_{j} (\Delta_{R})^{j}\,.
\end{equation}
We then expand 
the anomalous dimension $\gamma_{\rm IR}$ as 
\begin{align}
\label{eq:gamma_with_delta_Nf}
\gamma_{\rm IR}(\Delta_{R})=\sum_{i=1}^{\infty}k_{i}\left(\frac{\alpha_{\rm IR}}{4\pi}\right)^{i}
=\sum_{i=1}^{\infty}k_{i}\left(\sum_{j=1}^{\infty}a_{j}\Delta_{R}^{j}\right)^{i}.
\end{align} 
Similarly, the beta function, which vanishes at the IR fixed-point, is expanded as
\begin{align}
\beta_{\rm IR}(\Delta_{R})&
=-8\pi\sum_{i=1}^{\infty}b_{i}\left(\frac{\alpha_{\rm IR}}{4\pi}\right)^{i+1}
=-8\pi\sum_{i=1}^{\infty}b_{i}\left(\sum_{j=1}^{\infty}a_{j}\Delta_{R}^{j}\right)^{i+1}\nonumber\\
&=\sum_{i=2}^{\infty} d_{i}\Delta_{R}^{i}=0.
\label{eq:beta_IR}
\end{align}
Since the coefficients of the beta function $b_i$ depend on $N_R$, we need to expand them in powers of $\Delta_{R}$ to find $d_i$'s:
\begin{equation}
b_i(\Delta_{R})=
\sum_{n=0}^{\infty}\frac{1}{n!}(-1)^n \left.\frac{\partial^nb_i}{\partial N_R^n}\right|_{N_R=N_R^{\rm AF}}\Delta_{R}^n\,.
\end{equation}
As  $\Delta_{R}$ is an arbitrary positive number less than $N_{R}^{\rm AF}$, the coefficients $d_i$'s can be read off to find 
\begin{align}
d_{2}&=0, \nonumber \\
d_{3}&=-8\pi\left.\left[-a_{1}^{2}\frac{\partial b_{1}}{\partial N_{R}} +a_{1}^{3}b_{2}\right]\right|_{N_{R}=N_{R}^{\rm AF}},\nonumber\\
d_{4}&=-8\pi\left.\left[-2a_{1}a_{2}\frac{\partial b_{1}}{\partial N_{R}}-a_{1}^{3}\frac{\partial b_{2}}{\partial N_{R}}+3a_{1}^{2}a_{2}b_{2}+a_{1}^{4}b_{3}\right]\right|_{N_{R}=N_{R}^{\rm AF}}, \nonumber \\
d_{5}&=-8\pi\left[-\left(2a_{1}a_{3}+a_{2}^{2}\right)\frac{\partial b_{1}}{\partial N_{R}}+3\left(a_{1}a_{2}^{2}+a_{1}^{2}a_{3}\right)b_{2}\right.\nonumber\\
&\quad\qquad\left.\left.-3a_{1}^{2}a_{2}\frac{\partial b_{2}}{\partial N_{R}}-a_{1}^{4}\frac{\partial b_{3}}{\partial N_{R}}+
4a_{1}^{3}a_{2}b_{3}+a_{1}^{5}b_{4}\right]\right|_{N_{R}=N_{R}^{\rm AF}}\,, \nonumber \\
&\cdots, 
\label{eq:d_coeff}
\end{align}
where we use the properties that $b_{1}$ and $b_{2}$ have the terms proportional to a constant and $N_{R}$ only (see Eqs.~\ref{eq:b1} and \ref{eq:b2}), 
and the one-loop coefficient $b_{1}$ is zero at $N_{R}=N_{R}^{\rm AF}$. Now, since the beta function at the IR fixed-point 
vanishes for any $\Delta_{R}$, all $d_{i}$'s are identically zero 
to give the coefficents $a_i$'s of $\alpha_{\rm IR}$ as following:
\begin{align}
a_{1}&=\left.\frac{1}{b_{2}}\frac{\partial b_{1}}{\partial N_{R}}\right|_{N_{R}=N_{R}^{\rm AF}}, \nonumber\\
a_{2}
&=\left.\frac{1}{b_{2}^{3}}\left(\frac{\partial b_{1}}{\partial N_{R}}\right)\left[b_{2}\frac{\partial b_{2}}{\partial N_{R}}-\frac{\partial b_{1}}{\partial N_{R}}b_{3}\right]\right|_{N_{R}=N_{R}^{\rm AF}}, \nonumber\\
a_{3}&=\frac{1}{b_{2}^{5}}\frac{\partial b_{1}}{\partial N_{R}}\left[b_{2}^{2}\left(\frac{\partial b_{2}}{\partial N_{R}}\right)^{2}+2b_{3}^{2}\left(\frac{\partial b_{1}}{\partial N_{R}}\right)^{2}-3b_{2}b_{3}\frac{\partial b_{1}}{\partial N_{R}}\frac{\partial b_{2}}{\partial N_{R}}\right.\nonumber\\
&\quad\qquad\qquad\left.\left.+b_{2}^{2}\frac{\partial b_{1}}{\partial N_{R}}\frac{\partial b_{3}}{\partial N_{R}}-b_{2}b_{4}
\left(\frac{\partial b_{1}}{\partial N_{R}}\right)^{2}\right]\right|_{N_{R}=N_{R}^{\rm AF}},\nonumber \\
&\cdots.
\end{align}
Once $a_i$'s are given, we readily determine from Eq.~\ref{eq:gamma_with_delta_Nf} the coefficients $c_i$'s in the scheme-independent expansion of the anomalous dimension 
to find  
\begin{align}
c_{1}&=\left.a_{1}k_{1}\right|_{N_{R}=N_{R}^{\rm AF}},\nonumber\\
c_{2}&=\left.\left[a_{2}k_{1}+a_{1}^{2}k_{2}\right]\right|_{N_{R}=N_{R}^{\rm AF}},\nonumber\\
c_{3}&=\left.\left[a_{3}k_{1}+2a_{1}a_{2}k_{2}-a_{1}^{2}\frac{\partial k_{2}}{\partial N_{R}}+a_{1}^{3}k_{3}\right]\right|_{N_{R}=N_{R}^{\rm AF}}, \nonumber \\
&\cdots ,
\end{align}
where we used the fact that $k_1$ is a constant or $\frac{\partial k_{1}}{\partial N_{R}}=0$.

\subsection{Scheme (in)dependence of the critical condition $\gamma$CC}
\label{sec:SICC}

As discussed in section \ref{sec:CC}, two different forms of the critical condition, 
$\gamma_{\rm IR}=1$ and $\gamma_{\rm IR}(2-\gamma_{\rm IR})=1$, should agree with each other 
in full theory. However, they may differ and give different conformal windows, 
if truncated at a finite order in the perturbative expansion. 
Furthermore, if the anomalous dimension is evaluated in the expansion of the gauge coupling, it does depend on the renormalization scheme in general
and so does the truncated critical-condition. However, by using the scheme-independent series expansion of $\gamma_{\rm IR}$ discussed in the previous section, 
we could avoid the scheme-dependency to obtain more physical critical-conditions. 
Below, we discuss these issues in an exemplified case of $SU(3)$ gauge theory with $N_R$ Dirac fermions in the fundamental representation. 

To see this we first consider the scheme-dependent loop-expansion of the anomalous dimension up to the $4$-loop,
\beq
\gamma_{\rm IR}^{(\ell)}(\alpha_{\rm IR}^{(\ell)}) = \sum_{i=1}^{\ell}{k_{i}\left(\frac{\alpha_{\rm IR}^{(\ell)}}{4\pi}\right)^{i}}, 
\eeq
where $\ell=1,\,2,\,3$ and $4$. 
The coefficient at one-loop order, $k_1=6 C_2(R)$, is scheme independent. 
The coefficient $k_i$, as well as $b_i$ in \Eq{beta_IR}, 
have been computed in several different schemes such as the modified minimal subtraction ($\overline{\rm MS}$) scheme \cite{Chetyrkin:1997dh,Vermaseren:1997fq}, 
the modified regularization invariant ($\rm RI'$) scheme \cite{Gracey:2003yr}, 
and the minimal momentum subtraction ($\rm mMOM$) scheme \cite{Gracey:2013sca}. (See also \cite{Ryttov:2014nda}.)
Note that we take the same order in the loop expansion for both the beta function and the anomalous dimension, 
where $\alpha_{\rm IR}^{(\ell)}$ is obtained by equating the $\ell$-loop beta function to be zero. 
We then define the first critical condition at a given loop-order $\ell$ as $\gamma_{\rm IR}^{(\ell)} = 1$ 
for $2\leq \ell \leq 4$. 
Accordingly, the critical condition $\gamma_{\rm IR}(2-\gamma_{\rm IR})=1$ defines at each order as following:
\beq
2k_{1}\left(\frac{\alpha^{(2)}_{\rm IR}}{4\pi}\right)+(2k_{2}-k_{1}^{2})\left(\frac{\alpha^{(2)}_{\rm IR}}{4\pi}\right)^{2}=1,
\eeq
for the $2$-loop, 
\beq
2k_{1}\left(\frac{\alpha^{(3)}_{\rm IR}}{4\pi}\right)+(2k_{2}-k_{1}^{2})\left(\frac{\alpha^{(3)}_{\rm IR}}{4\pi}\right)^{2}
+(2k_{3}-2k_{1}k_{2})\left(\frac{\alpha^{(3)}_{\rm IR}}{4\pi}\right)^{3}=1,
\eeq
for the $3$-loop, and
\beq
2k_{1}\left(\frac{\alpha^{(4)}_{\rm IR}}{4\pi}\right)+(2k_{2}-k_{1}^{2})\left(\frac{\alpha^{(4)}_{\rm IR}}{4\pi}\right)^{2}
+(2k_{3}-2k_{1}k_{2})\left(\frac{\alpha^{(4)}_{\rm IR}}{4\pi}\right)^{3}
+(2k_{4}-2k_{1}k_{3}-k_{2}^{2})\left(\frac{\alpha^{(4)}_{\rm IR}}{4\pi}\right)^{4}=1,\nn \\
\eeq
for the $4$-loop. 

\begin{figure}
\centering
\includegraphics[width=0.49\textwidth]{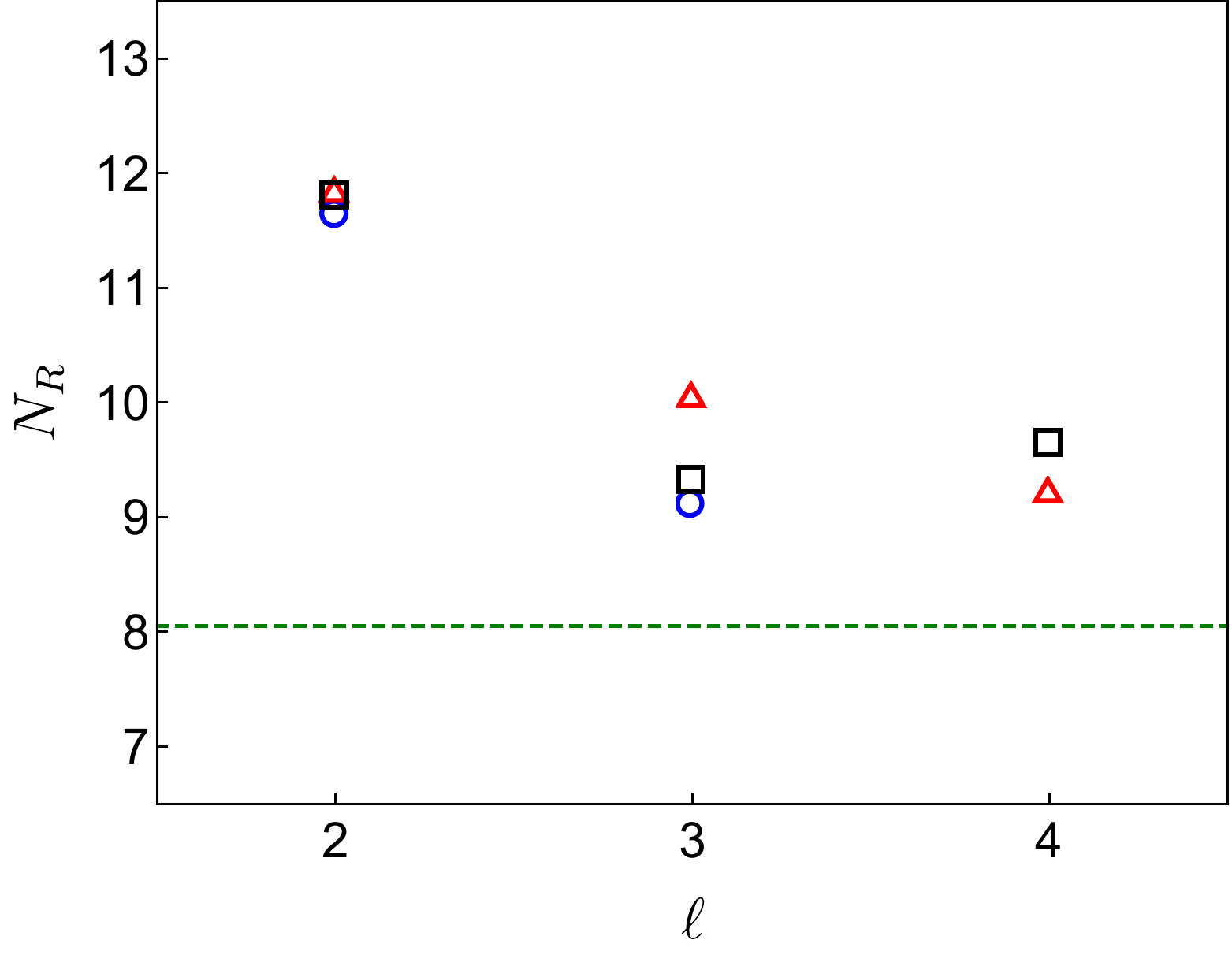}
\includegraphics[width=0.49\textwidth]{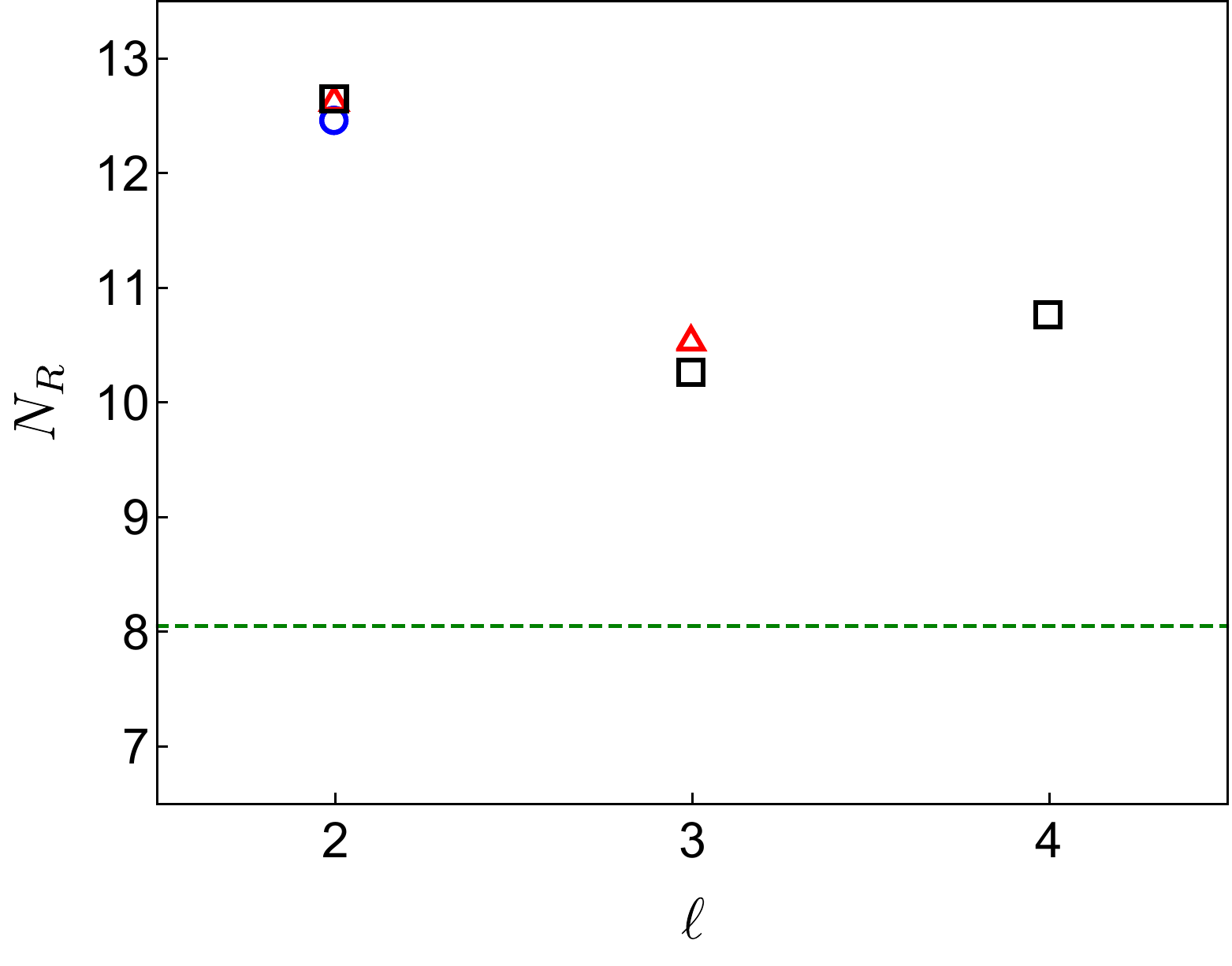}
\caption{The lower-bound of the conformal window in ${\rm SU}(3)$ gauge theory with $N_R$ fermions 
in the fundamental representation using the critical condition $\gamma$CC evaluated at $\ell$-th loop order. 
The green dashed line in both panels denotes the lower-bound from the $2$-loop beta function in \Eq{nc_2_loop}. 
In the left panel we present the lower bounds of the conformal window obtained from the critical condition $\gamma^{(\ell)}_{\rm IR}=1$, 
and the right panel we present the results from the condition $\gamma^{(\ell)}_{\rm IR}(2-\gamma^{(\ell)}_{\rm IR})=1$ with $2\leq \ell \leq 4$. 
The blue circle is for $\overline{\rm MS}$, the red triangle for $\rm RI'$, and the black square for $\rm mMOM$ schemes. 
}
\label{fig:su3_s_g1}
\end{figure}

Using the these critical conditions, 
we obtain the lower boundaries of the conformal window in the above three different schemes, 
$\overline{\rm MS}$, $\rm RI'$, and $\rm mMOM$, for the ${\rm SU}(3)$ gauge theory with $N_R$ Dirac fermions in the fundamental representation at $2$-loop, 
$3$-loop and $4$-loop orders, separately.
The results are shown in \Fig{su3_s_g1}. 
As seen in the figure, the two different critical-conditions give different results on the conformal window for each scheme. 
For comparison we also present the result obtained by the $2$-loop method in \Eq{nc_2_loop} by green dashed line. 
Note that in certain schemes  we could not find reasonable values of $N_R^c$ at the $3$- and $4$-loop orders: 
either the resulting values are below the $2$-loop value (green dashed line) or 
the solutions do not even exist. 
In the  $\rm mMOM$ scheme we find the lower bound $N_R^c$ up to the $4$-loop order, but we note 
the results from two different critical conditions do not converge to each other even if  we increase the loop order. 

Now, we consider the scheme-independent expression for the anomalous dimension and we truncate it at the $\ell$-th order in $\Delta_{R}$.  From the critical condition of $\gamma_{\rm IR}=1$, 
we take
\beq
\sum_{i=1}^{\ell}c_{i}(\Delta_{R})^{i}=1.
\label{eq:SICC_gamma}
\eeq
For the condition of $\gamma_{\rm IR}(2-\gamma_{\rm IR})=1$, we find
\beq
2c_{1}\Delta_R=1
\eeq
at the first order, 
\beq
2c_{1}\Delta_{R}+(2c_{2}-c_{1}^{2})\left(\Delta_{R}\right)^{2}=1
\eeq
at the second order, 
\beq
2c_{1}\Delta_{R}+(2c_{2}-c_{1}^{2})\left(\Delta_{R}\right)^{2}+(2c_{3}-2c_{1}c_{2})\left(\Delta_{R}\right)^{3}=1
\label{eq:SICC_gamma_3rd}
\eeq
at the third order, 
\beq
2c_{1}\Delta_{R}+(2c_{2}-c_{1}^{2})\left(\Delta_{R}\right)^{2}+(2c_{3}-2c_{1}c_{2})\left(\Delta_{R}\right)^{3} 
+(2c_{4}-2c_{1}c_{3}-c_{2}^{2})\left(\Delta_{R}\right)^{4}=1
\label{eq:SICC_gamma_4th}
\eeq
at the fourth order, etc. 
These conditions are clearly scheme-independent at each order, since the coefficients $c_i$'s 
and $\Delta_{R}$ are invariant under the change of schemes. 
Therefore, we can determine the physical and scheme-independent lower edge of the conformal window in the perturbation theory by expanding 
$\gamma_{\rm IR}$ and $\gamma_{\rm IR}(2-\gamma_{\rm IR})$ in powers of $\Delta_{R}$. 

\begin{figure}
\centering
\includegraphics[width=0.49\textwidth]{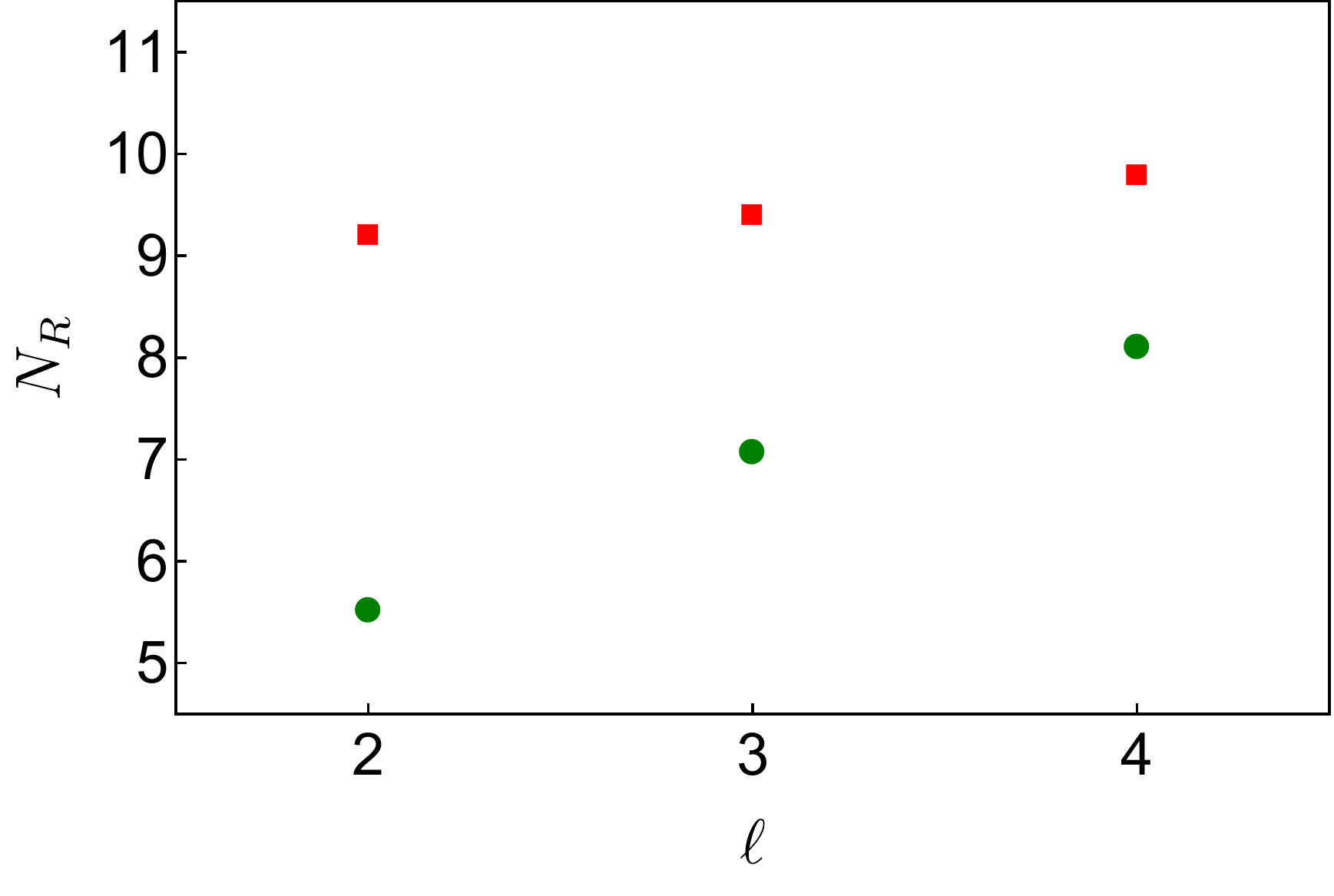}
\caption{Scheme-independent lower conformal-boundaries of ${\rm SU}(3)$ gauge theory with $N_R$ 
fundamental fermions calculated from the series expansion in $\Delta_{R}$ truncated at the order of $\ell=2,\,3,\,4$. 
The red square is from the condition 
$\gamma_{\rm IR}^{(\ell)}(2-\gamma_{\rm IR}^{(\ell)})=1$, while the green circle is from the $\gamma_{\rm IR}^{(\ell)}=1$ condition. 
}
\label{fig:su3_si}
\end{figure}

Using the values of the coefficient $c_i$ up to $i=4$, computed in Ref. \cite{Ryttov:2017kmx}, 
we determine the lower boundaries of the conformal window 
for ${\rm SU}(3)$ gauge theory with $N_R$ fundamental fermions at each order in $\Delta_{R}$ up to the fourth order. 
As seen in \Fig{su3_si}, the $\gamma_{\rm IR}^{(\ell)}(2-\gamma_{\rm IR}^{(\ell)})=1$ condition (red squares) yields much better convergence, compared to 
the $\gamma_{\rm IR}^{(\ell)}=1$ condition (green circles), 
and the resulting values are largely consistent 
with those evaluated from the scheme-dependent calculations at $3$rd and $4$th loop orders in \Fig{su3_s_g1}. 
We find that such a behavior persists in all the other theories considered in this work. 
We therefore use the scheme-independent critical condition $\gamma_{\rm IR}^{(\ell))} (2-\gamma_{\rm IR}^{(\ell)})=1$ 
for the definition of $\gamma$CC for the rest of this work. 

Of course these results should be taken with caution 
since the expansion parameter $\Delta_{R}$ becomes much bigger than unity near the lower edge of the conformal window 
and thus the convergence of the series expansion is not guaranteed in general. 
Nevertheless, compared to the results of the scheme-dependent expansions discussed above, 
the result in \Fig{su3_si} is promising, since it shows some evidence for the converegence: 
the resulting values of $N_R^c$ using two different critical conditions are getting closer to each other 
as we include the higher-order terms. 
Furthermore, the result obtained from the condition $\gamma_{\rm IR}(2-\gamma_{\rm IR})=1$
receives very small higher-order corrections. 
Note that $N_R^c$ determined by $\gamma_{\rm IR}^{(\ell)}=1$ monotonically increases as we increases the loop order $\ell$, 
which reflects the fact that the anomalous dimension $\gamma_{\rm IR}^{(\ell)}$ for a fixed $N_R$ monotonically increases with $\ell$ \cite{Ryttov:2017kmx}. 

\begin{table}
\begin{center}
\begin{tabular}{|c|c|c|c|c|c|c|c|}
\hline\hline
~~~{\bf G}~~~ & ~~~~$R$~~~~ &~2-loop($b_1=0$) & ~~SD~~ & BF($\gamma_{\rm IR}=2$) & 
~$\gamma$CC($\Delta_{R}^2$)~ & ~$\gamma$CC($\Delta_{R}^3$)~ & ~$\gamma$CC($\Delta_{R}^4$)~\\
\hline
$SU(2)$ & {\bf F} & $5.55$ & $8$ & $5.5$ & 
$5.69$ & $5.82$ & $6.22$ \\
$SU(3)$ & {\bf F} & $8.05$ & $11.2$ & $8.25$ & 
$9.2$ & $9.4$ & $9.8$  \\
$SU(2)$ & {\bf A} & $1.06$ & $2.08$ & $1.38$ & 
$1.86$ & $1.87$ & $1.92$ \\
$SU(3)$ & {\bf S2} & $1.22$ & $2.45$ & $1.65$ & 
$2.27$ & $2.29$ & $2.31$  \\
\hline\hline
\end{tabular}
\end{center}
\caption{%
\label{tab:sun_comparison}%
Comparison of the lower bounds of conformal window on the number of Dirac flavors, 
determined from various analytical approaches for $SU(2)$ and $SU(3)$ gauge groups: 2-loop denotes 
the two-loop beta function analysis, SD denotes the (traditional) Schwiner-Dyson analysis, 
BF denotes the unitarity bound on the all-orders beta function, 
and $\gamma$CC ($\Delta_{R}^\ell$) denotes the scheme-independent analysis of 
the critical anomalous dimension, expanded up to $\ell$-th order in perturbation theory. 
The fundamental, adjoint and two-index symmetric representations are denoted by {\bf F}, {\bf A} and {\bf S2}, respectively. 
}
\end{table}

We note that  our results for $N_R^c$ computed at $(\Delta_{R})^\ell$ for $\ell=2,\,3,\,4$ 
are placed below the value from SD but above those from the $2$-loop and BF methods. 
We present the resulting values in \Tab{sun_comparison}. 
We also report in the table the results for three other theories, corresponding to two-index symmetric (sextet) $SU(3)$, fundamental $SU(2)$,  and adjoint $SU(2)$. We find all of them show the similar trend. 
(We recall that $N_R$ denotes here the number of Dirac flavors.) 

Nonperturbative lattice results on the conformal window for the fundamental $SU(2)$ and $SU(3)$ theories are not conclusive yet. 
In the case of $SU(2)$, the lattice results indicate that $N_R=6$ is likely at the boundary of conformal window 
(e.g. see \cite{Leino:2017hgm} and references therein). 
For $SU(3)$, $N_R=12$ is likely to be inside the conformal window, though still controversal. 
$N_R=8$ is below the conformal window, but close enough to the conformal window so that it exhibits very different IR behaviors compared to QCD, 
and $N_R=10$ is largely unknown. 
(See \cite{Hasenfratz:2017qyr,Appelquist:2018yqe} and references therein.) 
These lattice results are more or less consistent with various analytical approaches except the SD method as shown in \Tab{sun_comparison}. 
In particular, the $\gamma$CC method predicts that $N_R=9$ or $10$ are likely 
near the boundary of the conformal window for the fundamental $SU(3)$ gauge theory and similarly $N_R=6$ for the fundamental $SU(2)$ gauge theory. 

In the cases of the adjoint $SU(2)$ and two-index symmetric $SU(3)$ gauge theories, 
the conformal window has also been estimated from several lattice calculations. 
The most recent results for the adjoint $SU(2)$ are summarized in Ref. \cite{Bergner:2017bky}, 
which shows that $N_R=1/2$ (supersymmetric Yang Mills) is confining, $N_R=2$ is IR conformal, 
and $N_R=3/2$ and $1$ are likely to be inside the conformal window. 
While various analytical estimates in \Tab{sun_comparison} are largely consistent with the lattice results, 
the $\gamma$CC results suggest that the critical $N_R^c$ is $\sim 2$ and thus $N_R=3/2$ and $1$ are rather in the broken phase 
(potentially near conformal). 
For the sextet $SU(3)$ theories $N_R=2$ has been extensively investigated by the means of lattice simulations 
with different types of discretization: with Wilson-type fermions the results are consistent with the theory being IR conformal, 
while with staggered fermions the results show near-conformal behaviors, 
see Refs.~\cite{Hansen:2017ejh,Fodor:2019vmw} and references therein. 
As shown in \Tab{sun_comparison}, $2$-loop and BF results support that $N_R=2$ sextet $SU(3)$ is IR conformal, 
but SD and $\gamma$CC results support that it is near-conformal.

\subsection{Scheme-independent critical conditions for multiple representations}
\label{sec:SICC_multirep}

As  explained in section \ref{sec:CW}, the upper-bound of the conformal window in multiple representations 
spans a hyper-surface of co-dimension one in the space of flavor numbers.  
For a two-representation case, widely used in the composite Higgs model, the pairs of flavor numbers ($N_\psi$, $N_\chi$) of its conformal window 
are bounded from above by ($N_\psi^{\rm AF}$, $N_\chi^{\rm AF}$) of fermions in the two different representations of $R_\psi$ and $R_\chi$. 
By the condition that the coefficient of the one-loop beta function, Eq.~\ref{eq:b1}, vanishes for the upper boundary of the conformal window
the pair of numbers ($N_\psi^{\rm AF}$, $N_\chi^{\rm AF}$) should satisfy 
\beq
4 N_{\psi}^{\rm AF} T(R_\psi) + 4 N_{\chi}^{\rm AF} T(R_\chi) = 11 C_{2}(G)\,,
\label{eq:AF_two_reps}
\eeq
which defines a set of points on a line in the space of representations $(R_\psi,R_\chi)$.  
To obtain the lower boundary of the conformal window for theories with two representations from the scheme-independent critical-condition, 
we first assume that at the IR fixed point the anomalous dimension of the bilinear operator 
of fermions in the representation $R_\chi$, $\gcir$, is larger than the one in the representation $R_\psi$, 
$\gpir$, 
so that the lower boundary of the conformal window is determined by 
$\gcir=1$ or 
$\gcir (2-\gcir)=1$. 

Since $C_2(G)$ is positive but finite, there exists a maximum value for $N_\psi^{\rm AF}\leq N_\psi^{\rm max}$.
It is then convenient to define $N_\chi^{\rm AF}$ for a fixed value of $N_\psi < N_\psi^{\rm max}$
\beq
N_{\chi}^{\rm AF}=\frac{11C_{2}(G)-4N_{\psi}T(\psi)}{4T(\chi)}.
\label{eq:AF_chi}
\eeq
Analogous to the case of a single representation, 
we also define $\Delta_{\chi} \equiv N_{\chi}^{\rm AF}-N_{\chi}$ 
and expand the anomalous dimension $\gcir$ in powers of $\Delta_{\chi}$
\beq
\gcir(\Delta_{\chi})=\sum_{i=1}^{\infty}{C_{i}(\chi,\psi)}(\Delta_{\chi})^{i}.
\label{eq:gamma_two_reps}
\eeq
The coefficients $C_i (\chi,\psi)$ have been computed to the $3$rd order in Ref.~\cite{Ryttov:2018uue} 
using the known pertubative results of the beta function and the anomalous dimension for the multiple representations, 
calculated up to the four-loop order in $\overline{\rm MS}$ scheme~\cite{Zoller:2016sgq}. 
At the finite order of the anomalous dimension, $\gcir^{(\ell)}$, 
we reuse the scheme-independent critical conditions in Eqs.~\ref{eq:SICC_gamma}-\ref{eq:SICC_gamma_3rd} 
by replacing the coefficients $c_i$ by $C_i (\chi,\psi)$ and $\Delta_{R}$ by $\Delta_{\chi}$, respectively, 
to determine the lower boundary of the conformal window. 
Note that both $C_i$ and $\Delta_{\chi}$ are functions of $N_\chi$ as well as $N_\psi$, 
and the critical conditions would yield the critical line of $(N_\psi^c,N_\chi^c)$. 

\section{Applications to two-representation composite Higgs models}
\label{sec:CHM}

We now turn our attention to the determination of conformal windows in 4-dimensional gauge theories 
with fermion matter fields in the two distinct representations 
relevant to composite Higgs and partial compositeness. 
The wish list of the underlying gauge models was first proposed in \cite{Ferretti:2013kya} 
and further refined in \cite{Ferretti:2016upr,Belyaev:2016ftv}, 
resulting in the most promising 12 models. 
Some of these models share the same gauge group and the same representations, 
but the details of the symmetry breaking patterns and/or the charge assignment under the non-anomalous $U(1)$ symmetry 
are different. 
Since we are interested in the possible extension of these models towards the conformal window, 
we rather classify them according to the gauge group: 
$SO(7)$, $SO(9)$ with fermions in the real fundamental and spinorial representations, 
$SO(11)$ with fermions in the real fundamental and pseudo-real spinorial representations, 
$SO(10)$ with fermions in the real fundamental and complex (chiral) spinorial representations, 
$Sp(4)$ with fermions in the pseudo-real fundamental and real two-index antisymmetric representations, 
$SU(4)$ with fermions in the complex fundamental and real two-index antisymmetric representations, 
and $SU(5)$ with fermions in the complex fundamental and two-index antisymmetric representations.
In Ref. \cite{Cacciapaglia:2019dsq} another type of UV complete composite Higgs models with fermion partial compositeness 
based on $Sp(4)$ gauge theories with $6$ antisymmetric and $12$ fundamental Weyl flavors were considered, 
which will be denoted by CVZ in this work. 
Note that throughout this section and \Appendix{loops} we denote $N_R$ for the number of Weyl spinors if the representation is real or pseudoreal, 
and for the number of Dirac flavors if the representation is complex. 

In phenomenological two-representation composite Higgs models the global symmetries are spontaneously broken 
by the fermion condensates at the scale of $\Lambda_\chi$, 
where part of pNGBs are identified as SM-like complex Higgs doublets. 
However, a partial compositeness prefers the gauge theories to be either conformal or near-conformal 
such that the baryonic operators and the SM quarks are linearly coupled for a wide range of energy scale, 
between the chiral symmetry breaking scale $\Lambda_\chi$ and the electroweak scale, $\Lambda_{\rm ew}$. 
This situation can simply be realized by introducing additional fermions which decouple just above $\Lambda_\chi$, 
and the extended gauge models eventually fall into the chirally broken phase 
with the expected symmetry breakings of the original models. 
Although in principle the scaling dimension of the baryonic operators can take any value between the classical dimension of $9/2$ 
and the unitary bound of $3/2$, 
the phenomenologically desired value for the top-partner is $\sim5/2$ so that the size of the linear coupling is the order of unity, $\mathcal{O}(1)$. 
Note that in this work we do not discuss the phenomenological aspects of near conformal dynamics for the composite Higgs and partial compositeness, 
but instead we map out the phase boundary of the conformal window which would be useful to provide a guidance 
for more dedicated nonpertubative studies on the IR dynamics. 

\begin{table}
\begin{center}
\begin{tabular}{|c|c|c|c|}
\hline\hline
Model & {\bf G} & ($R_\psi$, $R_\chi$) & ($N^{\rm min}_{\psi}$, $N^{\rm min}_{\chi}$) \\
\hline
M1, M3 & $SO(7)$ & {\bf (Sp, F)} & (5,13), (6,12), (7, 11), (8,10), (9,9), (10,8), (11,7), (12,6), (13,5)\\
M2, M4 & $SO(9)$ &  {\bf (F, Sp)} & (5,10), (7,9), (9,8), (11,7), (13,6), (15,5)\\
M5, M8, CVZ & $Sp(4)$ & {\bf (F, ${\rm \bf A_2}$)} & (4,9), (6,8), (8,7), (10,6), (12,5)\\
M6, M11 & $SU(4)$ &  {\bf ((F,${\rm \bf \overline{F}}$), ${\rm \bf A_2}$)} & (3,12), (4,11), (5,10), (6,9), (7,8), (8,7), (9,6), (10,5)\\
M7, M10 & $SO(10)$ & {\bf (F, (Sp,${\rm \bf \overline{Sp}}$))} & (5,6), (9,5), (13,4), (18,3)\\
M9 & $SO(11)$ & {\bf (F, Sp)} & (6,7), (10,6), (14,5), (18,4)\\
M12 & $SU(5)$ & {\bf ((F,${\rm \bf \overline{F}}$), ${\rm \bf (A_2,\overline{A}_2})$)} & (4,5), (7,4), (10,3)\\
\hline\hline
\end{tabular}
\end{center}
\caption{%
\label{tab:cw_two_reps_CHM}%
Pairs of minimum integer values for the numbers of flavors in two distinct representations for a given gauge group {\bf G} considered 
in the pNGB composite Higgs models to be in the conformal window.
Note that ${\bf F}$, ${\rm \bf A_2}$ and ${\rm \bf Sp}$ denote for the fundamental, 
the two-index antisymmetric and the spinorial irreducible representations, respectively, 
where a bar notation stands for the complex conjugate. 
In the first column we present the relevant models found in Refs. \cite{Belyaev:2016ftv, Cacciapaglia:2019dsq}.
}
\end{table}

In \Tab{cw_two_reps_CHM}, we summarize our findings on the pairs of minimal (integer) numbers $(N_\psi^{\rm min},N_\chi^{\rm min})$ 
for which the aforementioned two-representation gauge theories are in the conformal window. 
In other words, the theory falls into the chirally broken phase if we decrease any of  $N_\psi$ or $N_\chi$ by at least one 
from the values listed in the table. 
In the first column we also present the corresponding names of the models introduced in Ref. \cite{Belyaev:2016ftv}. 
Following the discussion in \Sec{CC}, we determine the phase boundary of the conformal transition 
when any of the representations reaches the critical condition. 
Let us denote by $R_\chi$ the representation that determines the conformal transition in accord with our notations in \Sec{SICC_multirep}, 
and the other representation by $R_\psi$. 
Note that the higher representation typically yields the larger value for the anomalous dimension, 
where some exceptions will be found for the small number of colors such as the case of $SO(7)$. 
We will come back to this issue later in this section. 
As we discussed in the previous section, our choice for $\gamma$CC is $\gcir (2-\gcir)=1$, 
which provides better convergence than $\gcir=1$ for all the considered cases of two representations. 
Since we only consider an extension of the partial composite Higgs models, 
we exclude the cases of which either of $N_\psi$ and $N_\chi$ is smaller than any of the values considered in the original models. 
We found that with these restrictions on $(N_\psi,N_\chi)$ the effective theories below 
which $N_\chi$ fermions are integrated out always develop a non-zero fermion condensate of $\psi$, 
i.e. $\Gpir > 1$, and thus no question arises to use the $\gamma$CC on $\gcir$ for the determination of the conformal window. 
Similar conclusion is obtained for the traditional Schwinger-Dyson approach in \Sec{SD}, 
where we use the critical coupling $\alpha_c$ for $R_\chi$ which is smaller than that for $R_\psi$. 

\begin{figure}
\begin{center}
\includegraphics[width=.49\textwidth]{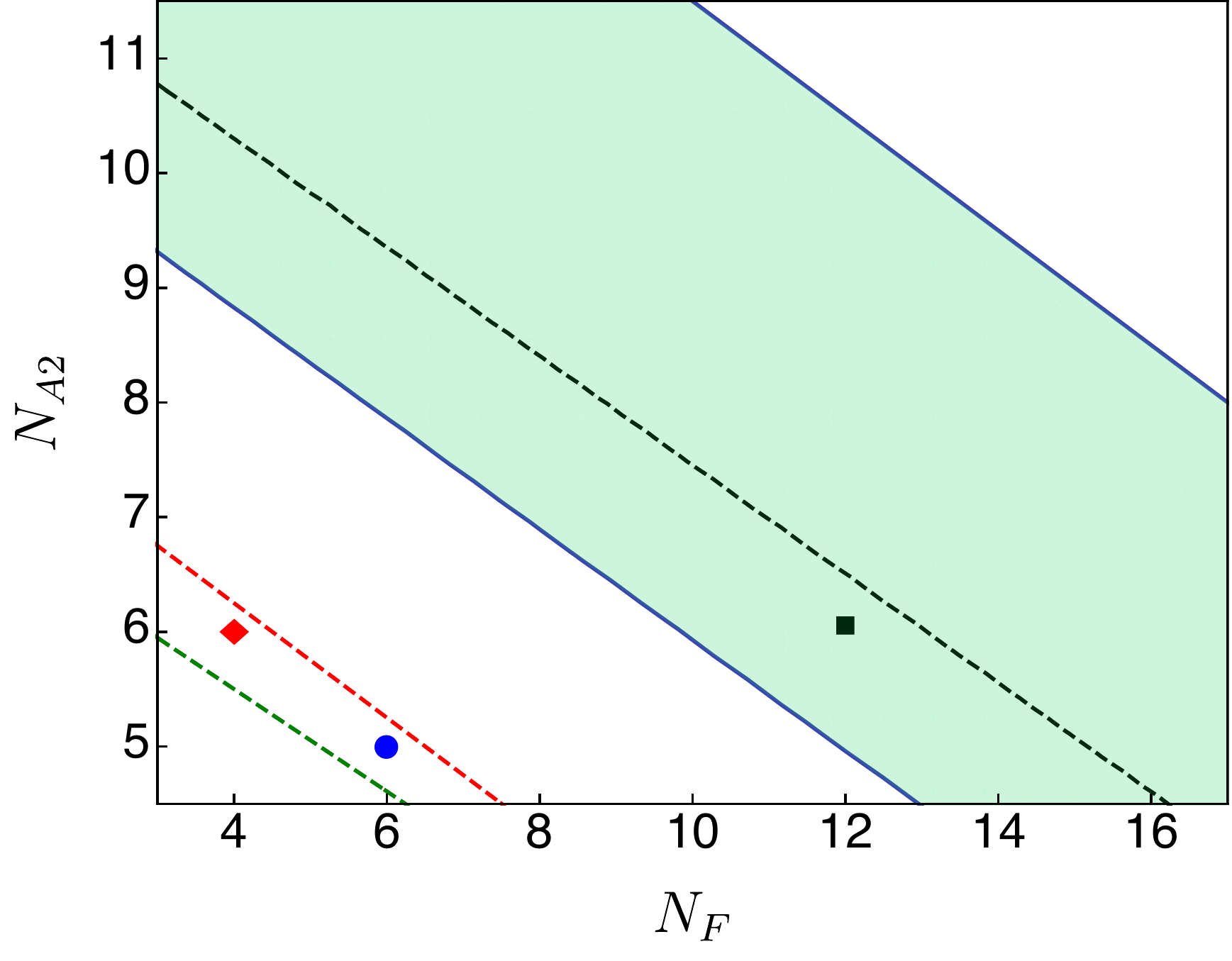}
\caption{%
\label{fig:sp4_cw}%
The estimated conformal window for $Sp(4)$ gauge theories 
with $N_{\rm F}$ fundamental and $N_{\rm A2}$ antisymmetric fermions. 
The upper bound of the shaded region is associated with the lost of asymptotic freedom, 
while the lower bound is determined by the critical condition $\gamma$CC. 
For comparison we also present the lower bounds of the conformal window estimated by other analytical methods: 
black, red and green dashed lines are for SD, BF, and $2$-loop results, respectively. 
M5, M8 and CVZ models are denoted by blue circle, red diamond and black square, respectively. 
}
\end{center}
\end{figure}

In gauge theories with two different representations the scheme-independent calculations of the anomalous dimension 
of fermion bilinears 
at a conformal IR fixed point are known to the cubic order in $\Delta_{\chi} = N_\chi^{\rm AF} - N_\chi$ 
with $N_\chi^{\rm AF}$ given in \Eq{AF_chi} \cite{Ryttov:2018uue}. 
The coefficients in the scheme-independent series expansions are functions of group invariants 
as well as the numbers of flavors in both the representations, $N_\chi$ and $N_\psi$. 
In \Appendix{group_inv} we present some relevant group theoretical quantities. 
Here, we show the results of the critical numbers of flavors obtained by applying $\gamma$CC to 
the highest order of $(\Delta_{\chi})^3$. 
The results obtained at lower orders of $\Delta_{\chi}$ and $(\Delta_{\chi})^2$ are shown in \Appendix{loops}. 

\begin{figure}
\begin{center}
\includegraphics[width=.49\textwidth]{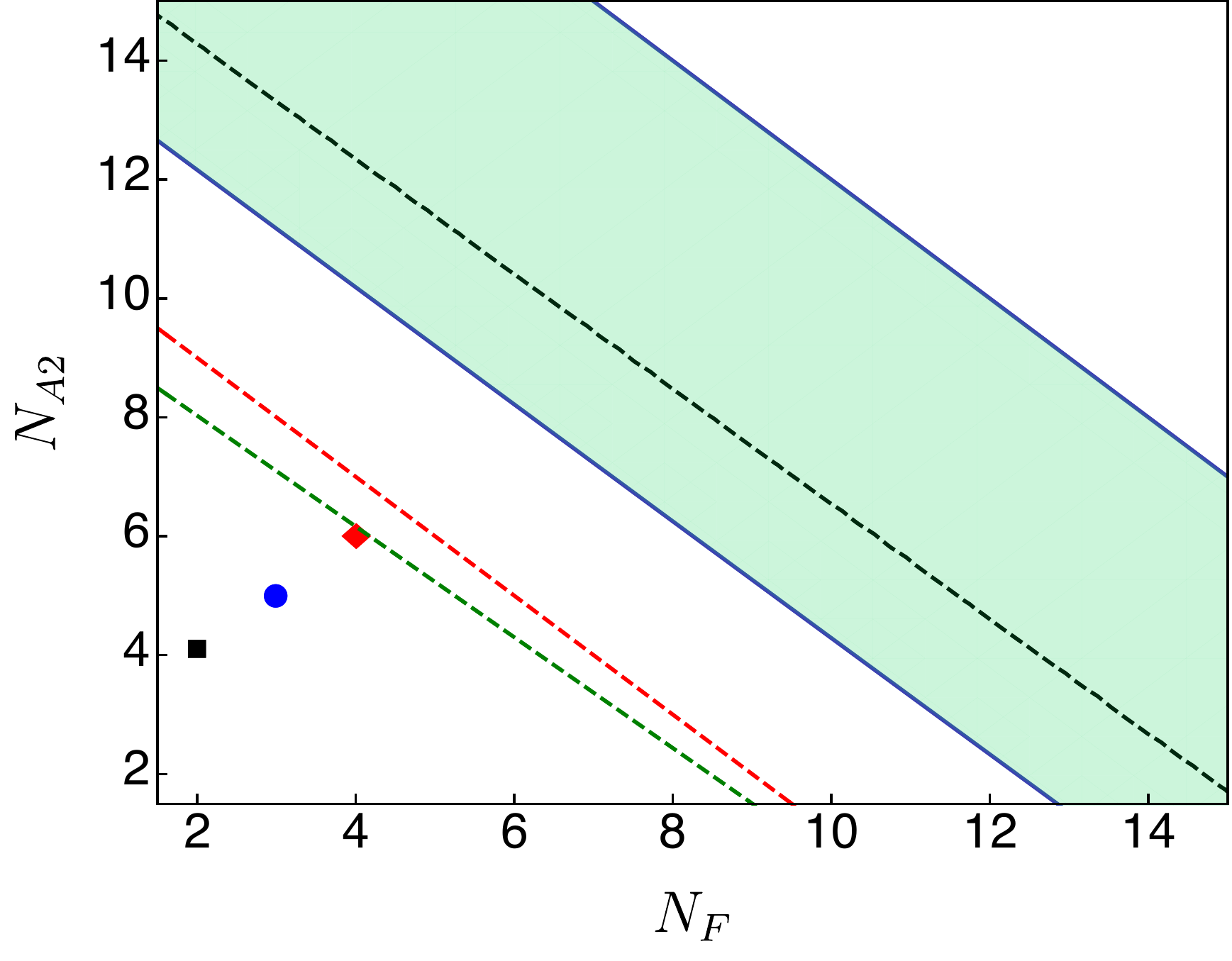}
\includegraphics[width=.49\textwidth]{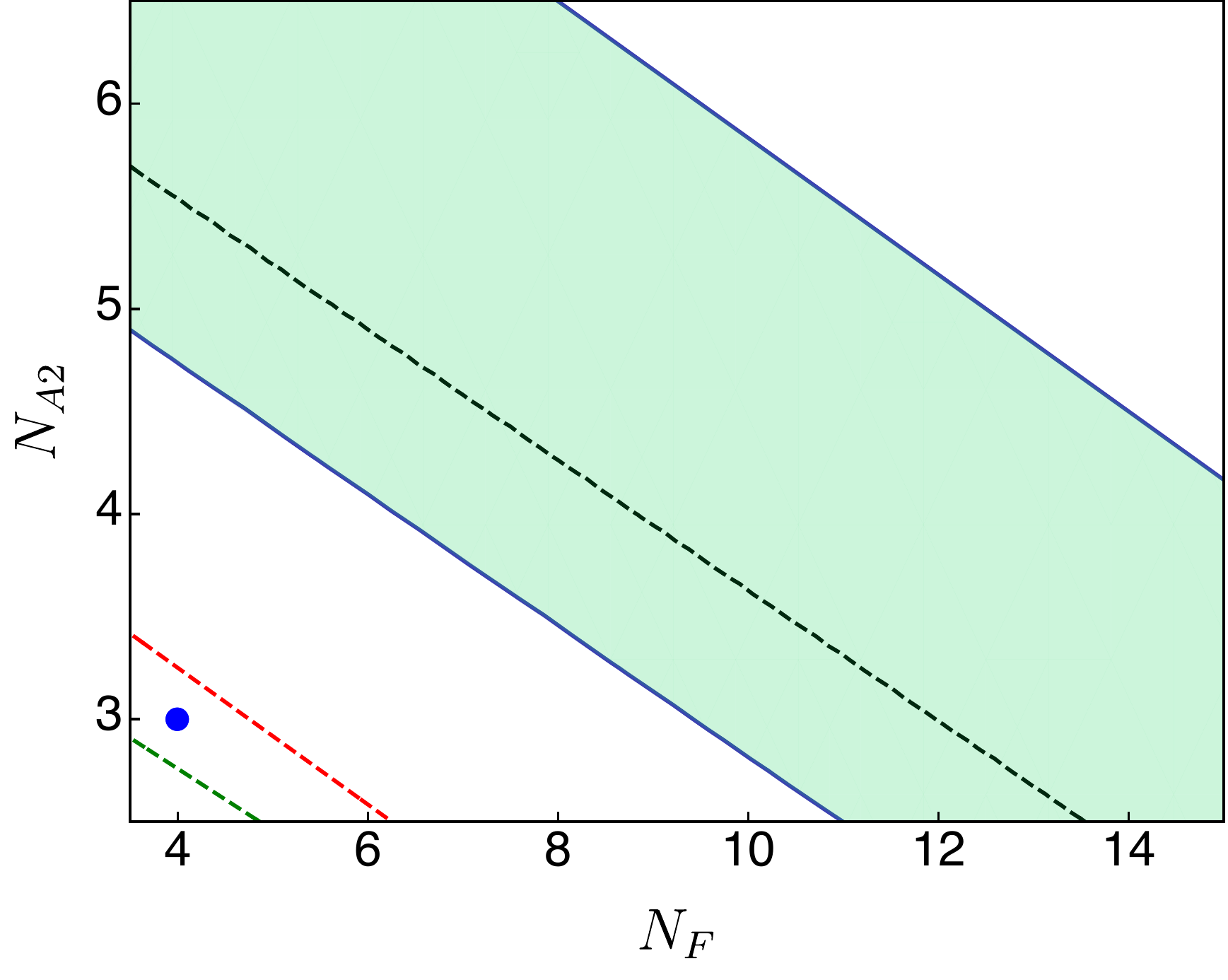}
\caption{%
\label{fig:su4_cw}%
The estimated conformal window in $SU(4)$ (left) and $SU(5)$ (right) gauge theories 
containing $N_{\rm F}$ fundamental and $N_{\rm A2}$ antisymmetric flavors. 
The upper bound of the shaded region is associated with the lost of asymptotic freedom, 
while the lower bound is determined by the critical condition $\gamma$CC. 
For comparison we also present the lower bounds of the conformal window estimated by other analytical methods: 
black, red and green dashed lines are for SD, BF, and $2$-loop results, respectively. 
M6 and M12 models are denoted by blue circles, M11 by red diamond, and the lattice $SU(4)$ model by black square. 
}
\end{center}
\end{figure}

\begin{figure}
\begin{center}
\includegraphics[width=.49\textwidth]{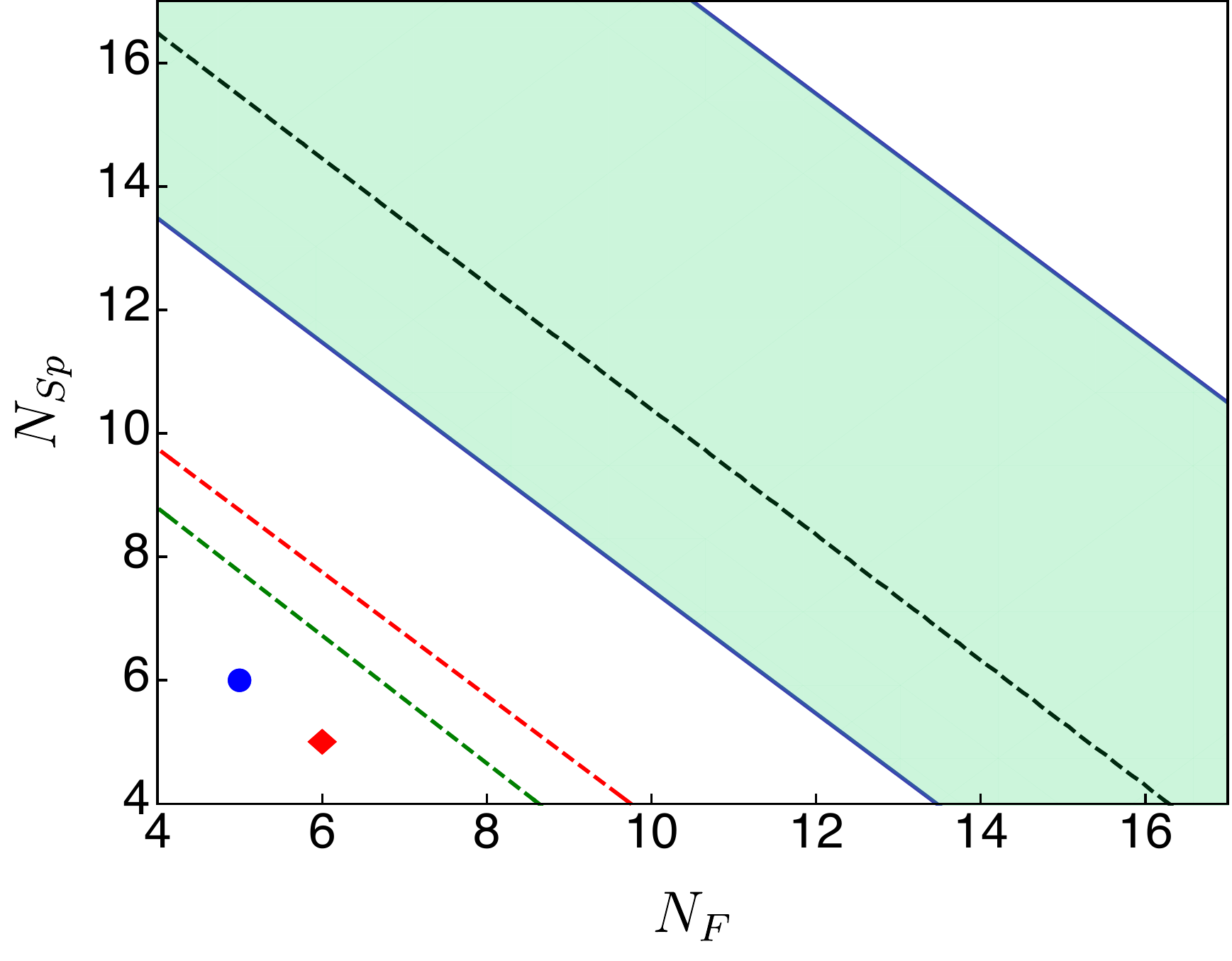}
\includegraphics[width=.49\textwidth]{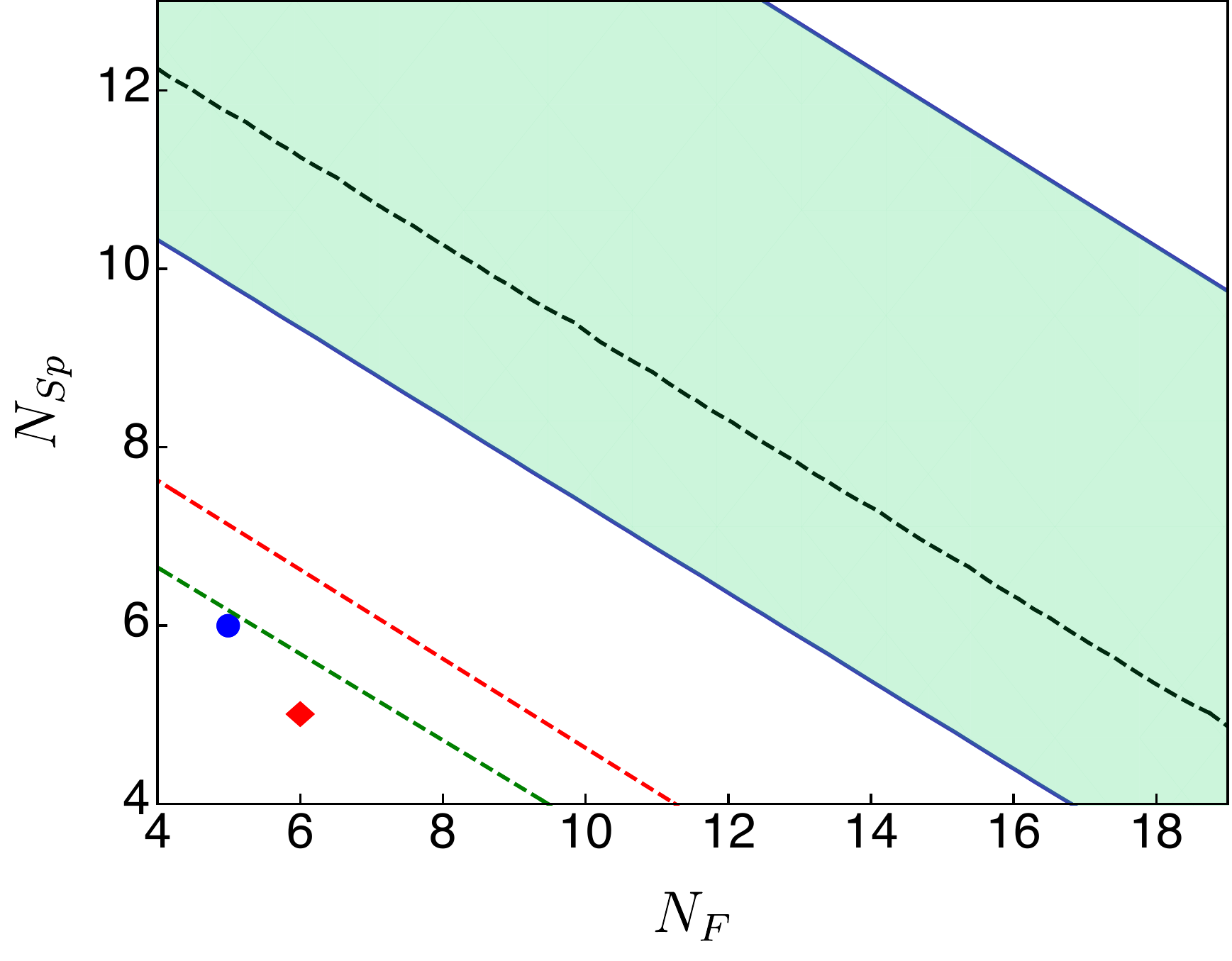}
\includegraphics[width=.49\textwidth]{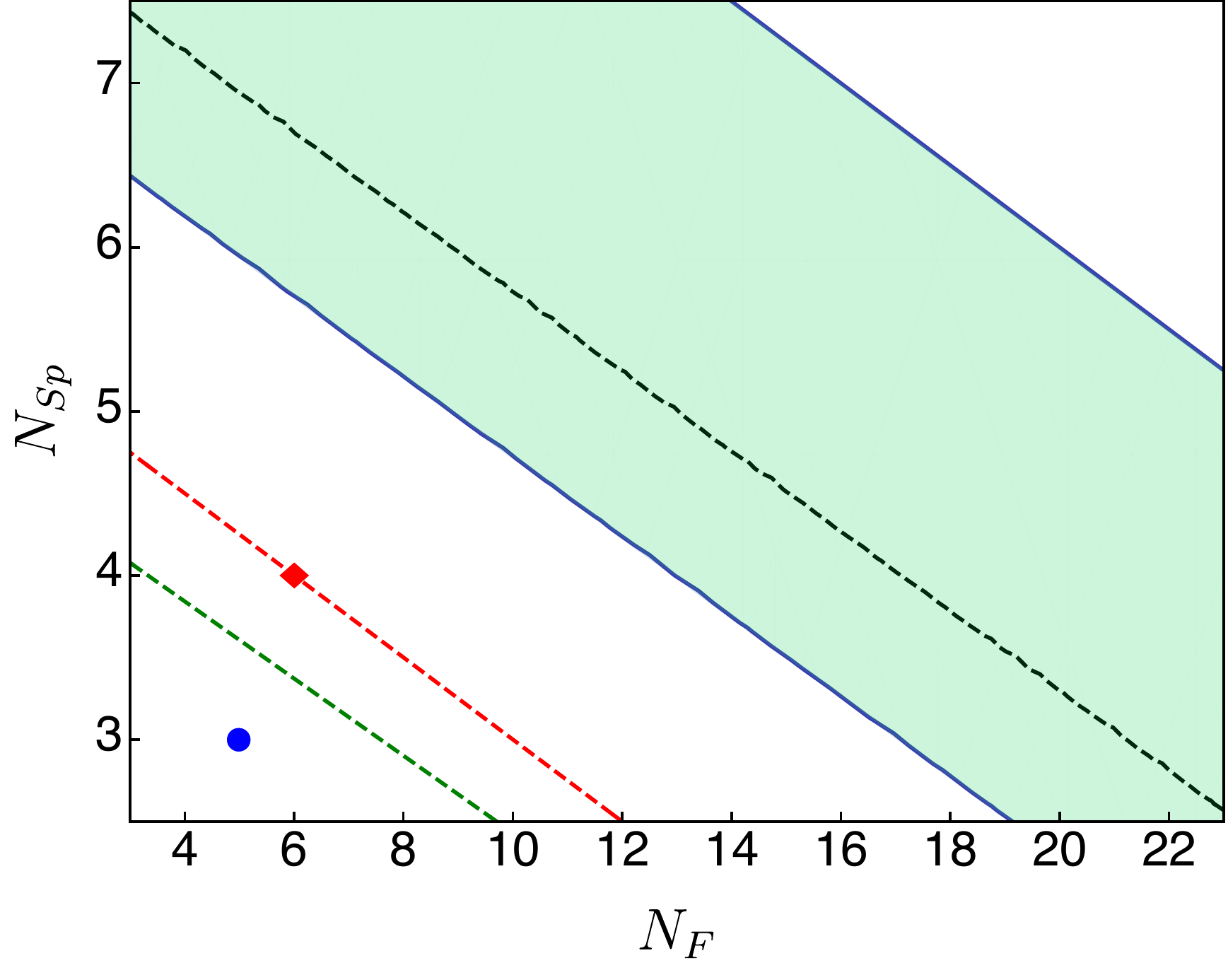}
\includegraphics[width=.49\textwidth]{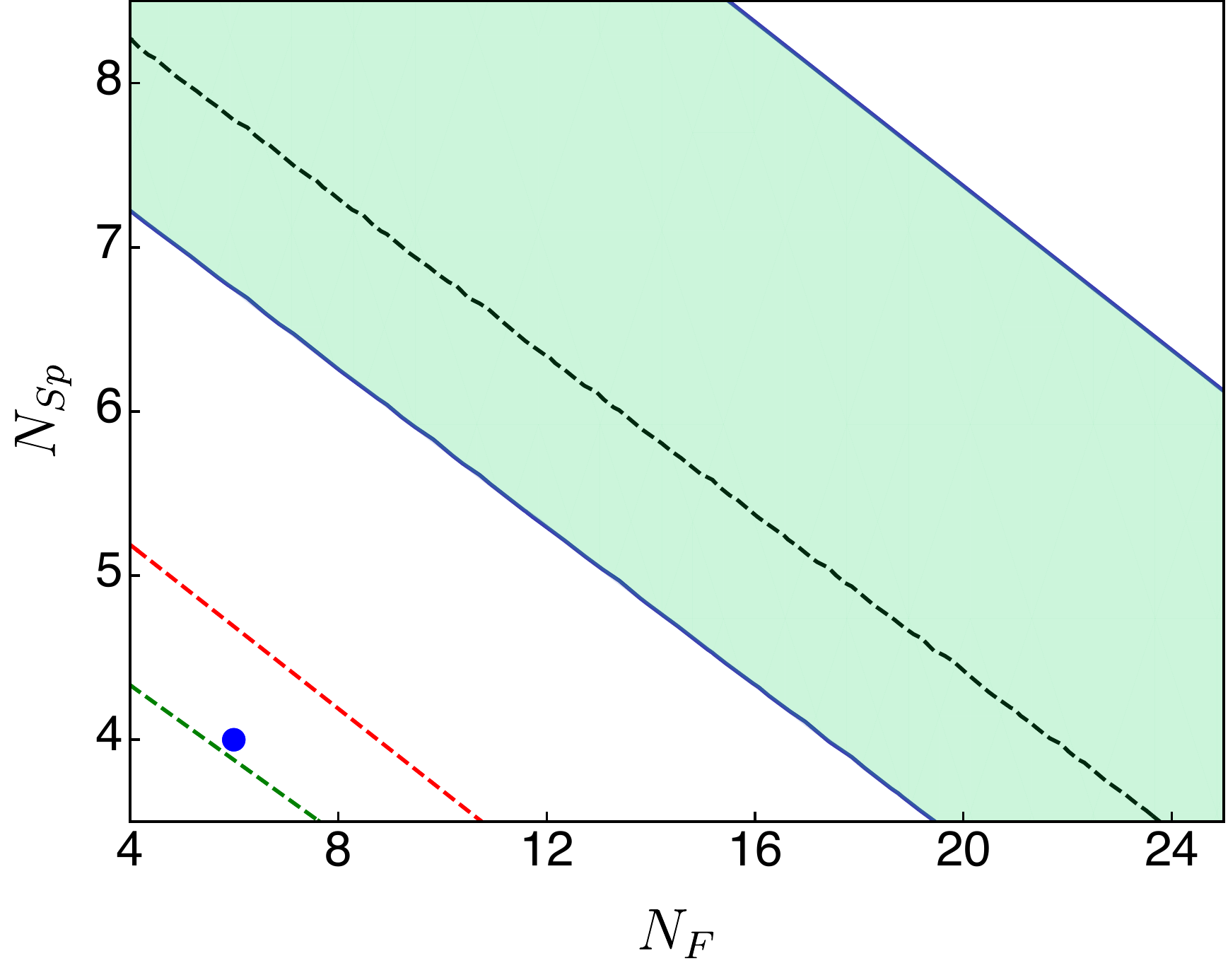}
\caption{%
\label{fig:so7_cw}%
The estimated conformal window in $SO(7)$ (top-left), $SO(9)$ (top-right), $SO(10)$ (bottom-left), 
$SO(11)$ (bottom-right) gauge theories 
containing $N_{\rm F}$ fundamental and $N_{\rm Sp}$ spinorial flavors. 
The upper bound of the shaded region is associated with the lost of asymptotic freedom, 
while the lower bound is determined by the critical condition $\gamma$CC. 
For comparison we also present the lower bounds of the conformal window estimated by other analytical methods: 
black, red and green dashed lines are for SD, BF, and $2$-loop results, respectively. 
M1, M2, M7, M9 models are denoted by blue circles, while M3, M4, M10 are denoted by red diamonds. 
}
\end{center}
\end{figure}

In Figs.~\ref{fig:sp4_cw}-\ref{fig:so7_cw} we present the map of the conformal window 
in the two-representation gauge theories of our interest.
The upper and lower bounds of the shaded region are obtained by using \Eq{AF_two_reps} 
and the critical condition $\gamma$CC, respectively. 
For comparison we also show the lower bounds estimated from the other analytical approaches, 
where green, red and black dashed lines are for $2$-loop, BF and SD methods, respectively. 
In the case of $Sp(4)$ gauge theories containing 
$N_{\rm F}$ fundamental and $N_{\rm A2}$ two-index antisymmetric fermions the results are shown in \Fig{sp4_cw}, 
where M5, M8 and CVZ models are denoted by circle, diamond and square shapes. 
The first two models are outside the conformal window, while the CZV model is slightly inside the conformal window. 
The model M8 has particularly received much attention since 
the corresponding lattice models are under investigation \cite{Bennett:2017kga,Lee:2018ztv,Bennett:2019cxd}.

In the left and right panels of \Fig{su4_cw}, we present the results for $SU(4)$ and $SU(5)$ gauge theories containing 
$N_{\rm F}$ fundamental and $N_{\rm A2}$ two-index antisymmetric fermions, respectively. 
Blue circles are for the models M6 and M12, while the red diamond is 
for the model M11 and the black square for the lattice $SU(4)$ model considered in Refs. 
\cite{Ayyar:2017qdf,Ayyar:2018zuk,Ayyar:2018ppa,Ayyar:2018glg,Ayyar:2019exp,Cossu:2019hse}. 
As seen in the figure, all the models are outside the conformal window. 
In particular, the lattice $SU(4)$ model which contains $N_{\rm F}=2$ Dirac fundamental and $N_{\rm A2}=4$ Weyl antisymmetric fermions 
is deep inside the chirally broken phase, 
which is consistent with the fact that numerical results showed the nonperturbative features of confinement and 
(spontaneous) global symmetry breaking \cite{Ayyar:2017qdf,Ayyar:2018ppa}.

In \Fig{so7_cw}, from left-top to right-bottom panels, 
we show the estimated conformal window for $SO(7)$, $SO(9)$, $SO(10)$ and $SO(11)$ gauge theories 
containing $N_{\rm F}$ fundamental and $N_{\rm Sp}$ spinorial representations. 
Note that, in contrast to other models, for $SO(7)$ we found that the dimension of the spinorial represenation 
is larger than that of the fundamental, but the anomalous dimension is smaller. 
Although we only learn this fact a posteriori since full IR dynamics 
is encoded in $\gpir$ and $\gcir$ in a complicated way, 
we can obtain some clues from the ladder approximation for multiple representations. 
As discussed in \Sec{SD}, if we depart from the conformal window 
the IR coupling $\alpha_{\rm IR}$ first reaches the critical coupling $\alpha_{\chi,\,c}$ 
for the representation having the largest value of the quadratic Casimir operator. 
In the case of $SO(7)$ we find that $C_2({\bf Sp}) < C_2({\bf F})$ so that the fundamental representation 
meets the critical condition first, 
while in the other cases $C_2({\bf F}) < C_2({\bf Sp})$. 
In \Fig{so7_cw}, blue circles denote the models M1, M2, M7, M9, 
and red diamonds denote the models M3, M4 and M10. 
All the models are outside the conformal window.

\section{Conclusion}
\label{sec:conclusion}

We have proposed an analytical method to determine the lower edge of the conformal window 
in a scheme-independent way by combining the conjectured critical condition on the anomalous dimension of 
a fermion bilinear, $\gamma_{\rm IR}=1$, which is responsible for the chiral phase transition, 
and the scheme-independent series expansion of $\gamma_{\rm IR}$ at a conformal IR fixed point 
with respect to $\Delta_{R}=(N_R^{\rm AF}-N_R)$. 
If all orders in the perturbative expansion are considered, 
this critical condition is identical to $\gamma_{\rm IR}(2-\gamma_{\rm IR})=1$, which is obtained from the Schwinger-Dyson analysis 
in the ladder approximation along with some working assumptions. 
However, at the finite order they yield different values for the critical number of flavors $N_R^c$ 
on the boundary of conformal and chirally broken phases. 
And it turns out that the latter condition shows much better convergence in the series expansion, 
while the resulting values obtained from both critical conditions approach to each other as we include higher order terms. 

In the illustrative examples of $SU(2)$ and $SU(3)$ gauge theories with 
Dirac fermions in various representations, 
we have determined $N_R^c$ using the scheme-independent critical condition 
on $\gamma_{\rm IR}$ up to $\mathcal{O}(\Delta_{R}^4)$, 
and compared to other analytical calculations. 
We find that the resulting values are larger than those estimated from the vanishing 2-loop coefficients and 
the all-order beta function with $\gamma_{\rm IR}=2$, but smaller than those from the traditional Schwinger-Dyson analysis in the ladder approximation. 
We also find that our values are largely consistent with the lattice results in the literature. 

We have extended the method of $\gamma$CC to the case of fermions in the $k$ different representations, 
where the critical $(k-1)$-dimensional surface is shown to be determined by the representation $R_\chi$ that reaches 
$\gcir=1$ first. 
Here, we assume that all the fermions in the representations other than $R_\chi$ eventually develop non-zero fermion condensates once  the fermions in $R_\chi$ decouple. 
We have applied this method to the gauge theories containing fermionic matter fields in the two distinct representations 
relevant to the models of composite Higgs and partial compositeness, and estimated the critical numbers of flavors ($N_\psi^c,\, N_\chi^c$) 
in the two dimensional space of $N_\psi$ and $N_\chi$ from the critical condition using $\gcir$ at the $3$rd order in $\Delta_{\chi}$. 
We find that all the partial composite Higgs models considered in Ref.~\cite{Belyaev:2016ftv} are in the chirally broken phase, 
while the CVZ model resides slightly inside the conformal window so that it is highly expected to have 
a large anomalous dimension of composite operators. 
While some of them are deep inside the broken phase (even below the $2$-loop estimation), 
models relatively close to the conformal window are such as M5, M8, M9, M10 and M12 models. 

In recent nonperturbative lattice studies  the $SU(3)$ gauge theory with $N_R=8$ fundamental fermions 
is shown to exhibit very different IR dynamics, having light scalar resonances, in contrast with QCD-like theories. 
Such findings may reflect the near-conformal dynamics. 
Note that, according to the analytical results presented in \Tab{sun_comparison}, 
the $8$-flavor $SU(3)$ model is slightly below all the estimates. 
Although we cannot simply generalize this specific case to generic multi-representation gauge theories, 
some of the partial composite Higgs models mentioned above can be good candidates for near-conformal theories. 
Hence, it would be encouraging to investigate such models in further details 
by means of nonperturbative lattice calculations.

\acknowledgments
The authors would like to thank G. Cacciapaglia, G. Ferretti, V. Leino and M. Piai for useful discussions. 
The work of BSK and JWL is supported in part by the National Research Foundation of Korea grant funded by the Korea government(MSIT) (NRF-2018R1C1B3001379). 
The work of DKH and JWL is supported in part by Korea Research Fellowship program funded by the Ministry of Science, ICT and Future Planning 
through the National Research Foundation of Korea (2016H1D3A1909283). 
The work of DKH is also supported in part by Basic Science Research Program through the National Research Foundation of Korea (NRF) 
funded by the Ministry of Education (NRF-2017R1D1A1B06033701).

\bibliography{cw}

\appendix

\section{Group invariants}
\label{sec:group_inv}

\begin{table}[t]
\begin{center}
\begin{tabular}{|c|c|c|c|c|}
\hline\hline
~SO(N)~ & ~~~~~~$d_{R}$~~~~~~ & ~~~~~~$T(R)$~~~~~~ & ~~~~~~$C_{2}(R)$~~~~~~ & 
~~~~~~$I_{4}(R)$~~~~~~\\
\hline
Fundamental & $N$ & $1$ & $\frac{N-1}{2}$ & $1$ \\
Chiral spinor (even N) & $2^{\frac{N-2}{2}}$ & $2^{\frac{N-8}{2}}$ & $\frac{N(N-1)}{16}$ & $-2^{\frac{N-10}{2}}$ \\
Real spinor (odd N) & $2^{\frac{N-1}{2}}$ & $2^{\frac{N-7}{2}}$ & $\frac{N(N-1)}{16}$ & $-2^{\frac{N-9}{2}}$ \\
Adjoint & $\frac{N(N-1)}{2}$ & $N-2$ & $N-2$ & $N-8$ \\
Rank-2 symmetric & $\frac{(N-1)(N+2)}{2}$ & $N+2$ & $N$ & $N+8$ \\
\hline\hline
\end{tabular}
\end{center}
\caption{%
\label{tab:so_group}%
Group invariants for various representations in $SO(N)$ gauge group.
}
\end{table}

\begin{table}[t]
\begin{center}
\begin{tabular}{|c|c|c|c|c|}
\hline\hline
~SU(N)~ & ~~~~~~$d_{R}$~~~~~~ & ~~~~~~$T(R)$~~~~~~ & ~~~~~~$C_{2}(R)$~~~~~~ & 
~~~~~~$I_{4}(R)$~~~~~~\\
\hline
Fundamental & $N$ & $\frac{1}{2}$ & $\frac{N^{2}-1}{2N}$ & $1$ \\
Adjoint & $N^{2}-1$ & $N$ & $N$ & $2N$ \\
Rank-2 symmetric & $\frac{N(N+1)}{2}$ & $\frac{N+2}{2}$ & $\frac{(N-1)(N+2)}{N}$ & $N+8$ \\
Rank-2 antisymmetric & $\frac{N(N-1)}{2}$ & $\frac{N-2}{2}$ & $\frac{(N+1)(N-2)}{N}$ & $N-8$ \\
\hline\hline
\end{tabular}
\end{center}
\caption{%
\label{tab:su_group}%
Group invariants for various represenatations in $SU(N)$ gauge group.
}
\end{table}

\begin{table}[t]
\begin{center}
\begin{tabular}{|c|c|c|c|c|}
\hline\hline
~Sp(N)~ & ~~~~~~$d_{R}$~~~~~~ & ~~~~~~$T(R)$~~~~~~ & ~~~~~~$C_{2}(R)$~~~~~~ & 
~~~~~~$I_{4}(R)$~~~~~~\\
\hline
Fundamental & $N$ & $\frac{1}{2}$ & $\frac{N+1}{4}$ & $1$ \\
Adjoint & $\frac{N(N+1)}{2}$ & $\frac{N+2}{2}$ & $\frac{N+2}{2}$ & $N+8$ \\
Rank-2 antisymmetric & $\frac{(N+1)(N-2)}{2}$ & $\frac{N-2}{2}$ & $\frac{N}{2}$ & $N-8$ \\
\hline\hline
\end{tabular}
\end{center}
\caption{%
\label{tab:sp_group}%
Group invariants for various representations in $Sp(N)$ gauge group.
}
\end{table}

\begin{table}[t]
\begin{center}
\begin{tabular}{|c|c|c|c|}
\hline\hline
~~ & ~SO(N)~ & ~SU(N)~ & ~Sp(N)~\\
\hline
$d^{abcd}d^{abcd}/d_{A}$ & $\frac{(d_{A}-1)(d_{A}-3)}{12(d_{A}+2)}$ & $\frac{(d_{A}-3)(d_{A}-8)}{96(d_{A}+2)}$ & $\frac{(d_{A}-1)(d_{A}-3)}{192(d_{A}+2)}$ \\
\hline\hline
\end{tabular}
\end{center}
\caption{%
\label{tab:product_d}%
Values of $d^{abcd}d^{abcd}/d_{A}$ for $SO(N)$, $SU(N)$, and $Sp(N)$ gauge groups.
}
\end{table}

\begin{table}[t]
\begin{center}
\begin{tabular}{|c|c|c|}
\hline\hline
~~ & ~SO(N) (even N)~ &~SO(N) (odd N)~\\
\hline
$d_{\bf F}^{abcd}d_{\bf F}^{abcd}/d_{A}$&$\frac{1}{24} \left(N^2-N+4\right)$&$\frac{1}{24} \left(N^2-N+4\right)$\\
$d_{\bf Sp}^{abcd}d_{\bf Sp}^{abcd}/d_{A}$& $\frac{1}{3} 2^{N-15} \left(13 N^2-61 N+76\right)$& $\frac{1}{3} 2^{N-14} \left(13 N^2-61 N+76\right)$ \\
$d_{\bf F}^{abcd}d_{\bf Sp}^{abcd}/d_{A}$& $-\frac{1}{3} 2^{\frac{N}{2}-8} \left(N^2-7 N+7\right)$& $-\frac{1}{3} 2^{\frac{N-15}{2}} \left(N^2-7 N+7\right)$ \\
$d_{\bf A}^{abcd}d_{\bf A}^{abcd}/d_{A}$&$\frac{1}{24} (N-2) \left(N^3-15 N^2+138 N-296\right)$&$\frac{1}{24} (N-2) \left(N^3-15 N^2+138 N-296\right)$\\
$d_{\bf F}^{abcd}d_{\bf A}^{abcd}/d_{A}$&$\frac{1}{24} (N-2) \left(N^2-7 N+22\right)$ &$\frac{1}{24} (N-2) \left(N^2-7 N+22\right)$ \\
$d_{\bf Sp}^{abcd}d_{\bf A}^{abcd}/d_{A}$&$-\frac{1}{3} 2^{\frac{N}{2}-8} \left(N^3-24 N^2+96 N-104\right)$&$-\frac{1}{3} 2^{\frac{N-15}{2}} \left(N^3-24 N^2+96 N-104\right)$ \\
\hline\hline
\end{tabular}
\end{center}
\caption{%
\label{tab:drdr_so}%
Values of $d_{R}^{abcd}d_{R'}^{abcd}/d_{A}$ in $SO(N)$ gauge groups with $N\geq3$.
We denote {\bf F}, {\bf Sp}, and {\bf A} for fundamental, spinor, adjoint representations. 
}
\end{table}

\begin{table}[t]
\begin{center}
\begin{tabular}{|c|c|c|}
\hline\hline
~~ & ~SU(N)~ &~Sp(N)~\\
\hline
$d_{\bf F}^{abcd}d_{\bf F}^{abcd}/d_{A}$&$\frac{N^4-6 N^2+18}{96 N^2}$&$\frac{1}{384}(N^2 + N + 4)$\\
$d_{\bf A2}^{abcd}d_{\bf A2}^{abcd}/d_{A}$& $\frac{(N-2) \left(N^5-14 N^4+72 N^3+48 N^2-288 N-576\right)}{96 N^2}$& $\frac{1}{384}(N - 2) (N^3 - 13 N^2 + 110 N - 104)$ \\
$d_{\bf F}^{abcd}d_{\bf A2}^{abcd}/d_{A}$& $\frac{N^5-8 N^4+6 N^3+48 N^2-144}{96 N^2}$& $\frac{1}{384} (-20 + 20 N - 7 N^2 + N^3)$ \\
$d_{\bf A}^{abcd}d_{\bf A}^{abcd}/d_{A}$&$\frac{1}{24} N^2 \left(N^2+36\right)$&$\frac{1}{384}(N + 2) (N^3 + 15 N^2 + 138 N + 296)$\\
$d_{\bf F}^{abcd}d_{\bf A}^{abcd}/d_{A}$&$\frac{1}{48} N \left(N^2+6\right)$ &$\frac{1}{384}(N + 2) (N^2 + 7 N + 22)$ \\
$d_{\bf A2}^{abcd}d_{\bf A}^{abcd}/d_{A}$&$\frac{1}{48} N (n-2) \left(N^2-6 N+24\right)$&$\frac{1}{384}(N + 2) (N - 2) (N^2 + N + 28)$ \\
\hline\hline
\end{tabular}
\end{center}
\caption{%
\label{tab:drdr_su}%
Values of $d_{R}^{abcd}d_{R'}^{abcd}/d_{A}$ in $SU(N)$ and $Sp(N)$ gauge groups with $N\geq2$.
We denote {\bf F}, {\bf A}, and {\bf A2} for fundamental, adjoint, and rank-2 antisymmetric representations. 
}
\end{table}

In this appendix, we summarize the group invariants needed to calculate the coefficients $C_\ell(R,R')$ 
in \Eq{gamma_two_reps} for the scheme-independent series expansions of $\gir(\Delta_{R})$ 
to $\ell=3$ in the cases of two different representations. 
As the coefficients for $\ell \geq 3$ involve four-loop results of the RG beta function, 
we need the group-invariant products of four-index quantities such as $d_{R}^{abcd} d_{R'}^{abcd}/d_{A}$ 
in addition to the trace normalization factor $T(R)$ ($T(R')$) 
and the eigenvalues of the quadratic Casimir operator $C_2(R)$ ($C_2(R')$). 
Here, we denote $A$ for the adjoint representation of gauge group $G$ and $d_A$ for its dimension. 
For the details of our discussion we refer the reader to Refs. \cite{Ryttov:2017dhd,Ryttov:2018uue} and references therein.

For a given representation $R$ the totally symmetric four-index quantity is defined as
\beq
d_{R}^{abcd}=\frac{1}{3!}{\rm Tr}\left[T^{a}\left(T^{b}T^{c}T^{d}+T^{b}T^{d}T^{c}
+T^{c}T^{b}T^{d}+T^{c}T^{d}T^{b}+T^{d}T^{b}T^{c}+T^{d}T^{c}T^{b}\right)\right],
\label{eq:dr_trace}
\eeq
where $T^a$ is the generators in $R$. 
In terms of group invariants this can be rewritten as
\beq
\label{eq:dR}
d_{R}^{abcd}=I_{4,R}\,d^{abcd}+\left(\frac{T(R)}{d_{A}+2}\right)\left(C_{2}(R)-\frac{1}{6}C_{2}(A)\right)
\left(\delta^{ab}\delta^{cd}+\delta^{ac}\delta^{bd}+\delta^{ad}\delta^{bc}\right).
\eeq
Here, $d^{abcd}$ is a traceless tensor, satisfying $\delta_{ab} d^{abcd}=0$, etc., 
which only depends on the group $G$. 
$I_{4,R}$ is a quartic group invariant. 
We list the values of the group invariants $d_A$, $T(R)$, $C_2(R)$, $I_{4,R}$ 
for the relevant representations in Tables \ref{tab:so_group}, \ref{tab:su_group} and \ref{tab:sp_group} 
for $SO(N)$, $SU(N)$ and $Sp(N)$, respectively.

Using the expression for $d_{R}^{abcd}$ in \Eq{dR}, one can obtain 
\beq
\frac{d_{R}^{abcd}d_{R'}^{abcd}}{d_{A}}&=&I_{4,R}\,I_{4,R'}\,\frac{d^{abcd}d^{abcd}}{d_{A}}\nn \\
&&+\left(\frac{3}{d_{A}+2}\right)T(R)\,T(R')
\left(C_{2}(R)-\frac{1}{6}C_{2}(A)\right)\left(C_{2}(R')-\frac{1}{6}C_{2}(A)\right).
\label{eq:drdr}
\eeq
The gauge invariant products of $d^{abcd}$ in the first term are independent on the representation. 
In \Tab{product_d} we present the results for $SO(N)$, $SU(N)$ and $Sp(N)$ gauge groups \cite{vanRitbergen:1997va}. 
Finally, we present the resulting values of 
$d_{R}^{abcd}d_{R'}^{abcd}/d_{A}$ for $SO(N)$, $SU(N)$, and $Sp(N)$ gauge groups 
in Tables \ref{tab:drdr_so} and \ref{tab:drdr_su}. 
Note that in the tables we only present the results of the two different representations relevant 
to the models for composite Higgs and partial compositeness. 
It is straightforward to obtain the results of other possibilities by using 
\Eq{drdr} and the group invariants presented in Tables \ref{tab:so_group}, \ref{tab:su_group}, \ref{tab:sp_group} and \ref{tab:product_d}. 
Part of the results are found in Ref. \cite{Ryttov:2017dhd}.

\section{Results on $\gamma$CC from lower orders in the scheme-independent series expansions}
\label{sec:loops}

\begin{figure}
\begin{center}
\includegraphics[width=.47\textwidth]{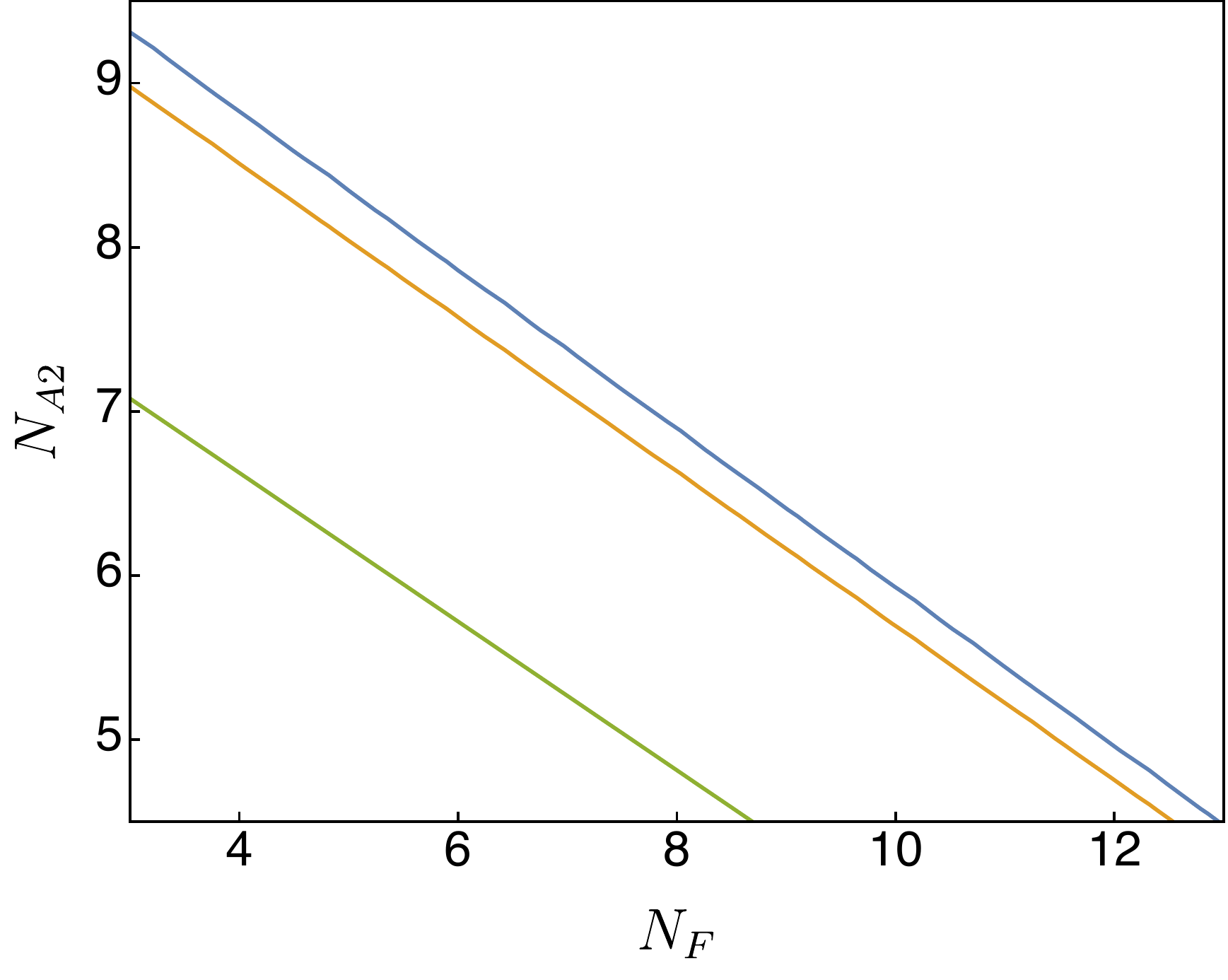}
\caption{%
\label{fig:sp4_loops}%
Estimation of the lower edge of the conformal window in $Sp(4)$ gauge theories 
containing $N_{\rm F}$ fundamental and $N_{\rm A2}$ antisymmetric flavors. 
We use the critical condition $\gamma$CC for the anomalous dimension of the representation $\chi$ at an IR fixed point, $\gcir (2-\gcir)=1$, 
and its scheme-independent series expansions truncated at $\Delta_\chi$ (green), $(\Delta_\chi)^2$ (yellow) and $(\Delta_\chi)^3$ (blue) orders, 
where we identify $R_\chi={\rm \bf A2}$. 
}
\end{center}
\end{figure}

\begin{figure}
\begin{center}
\includegraphics[width=.47\textwidth]{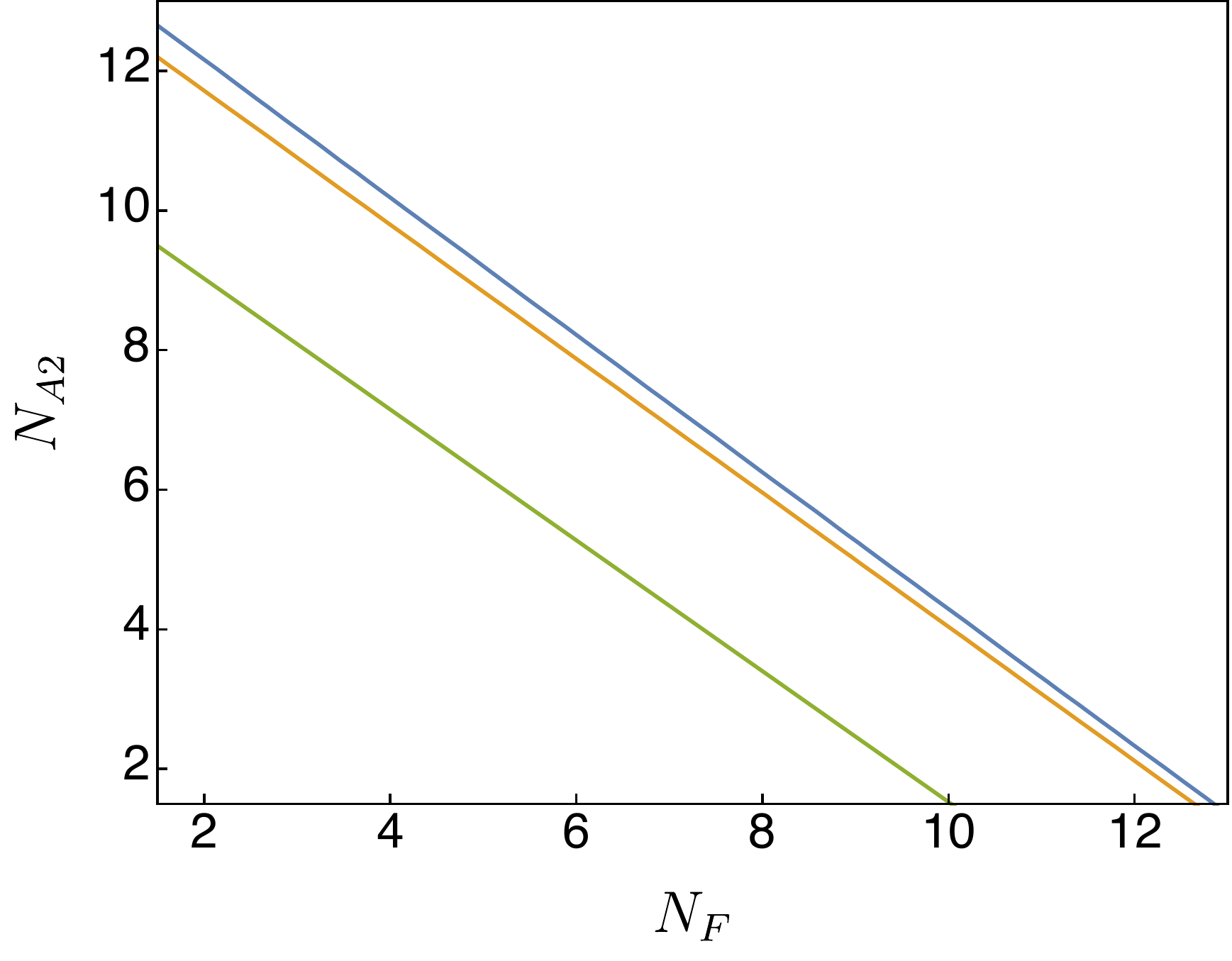}
\includegraphics[width=.47\textwidth]{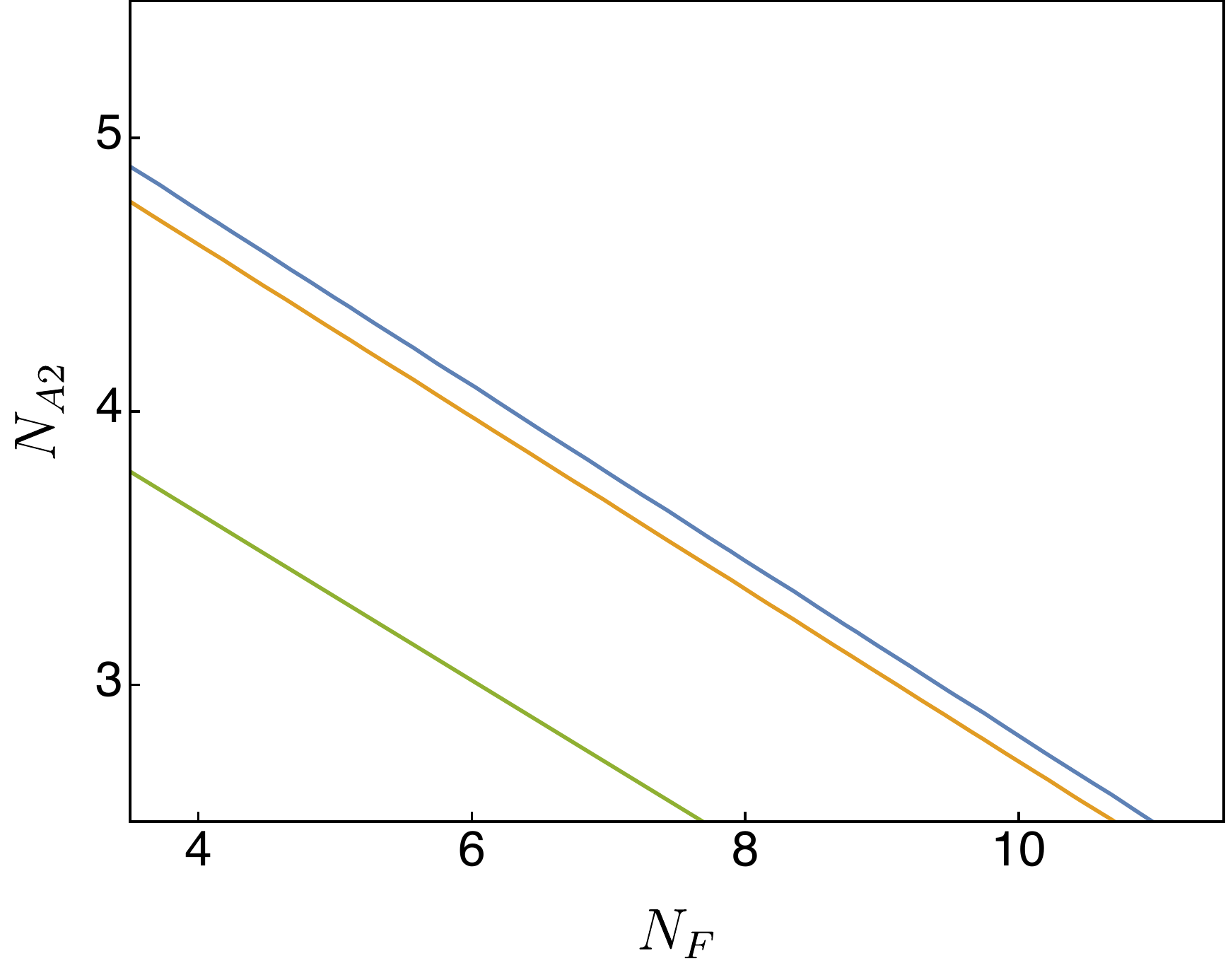}
\caption{%
\label{fig:su4_loops}%
Estimation of the lower edge of the conformal window in $SU(4)$ (left) and $SU(5)$ (right) gauge theories 
containing $N_{\rm F}$ fundamental and $N_{\rm A2}$ antisymmetric flavors. 
We use the critical condition $\gamma$CC for the anomalous dimension of the representation $\chi$ at an IR fixed point, $\gcir (2-\gcir)=1$, 
and its scheme-independent series expansions truncated at $\Delta_\chi$ (green), $(\Delta_\chi)^2$ (yellow) and $(\Delta_\chi)^3$ (blue) orders, 
where we identify $R_\chi={\rm \bf A2}$. 
}
\end{center}
\end{figure}

\begin{figure}
\begin{center}
\includegraphics[width=.47\textwidth]{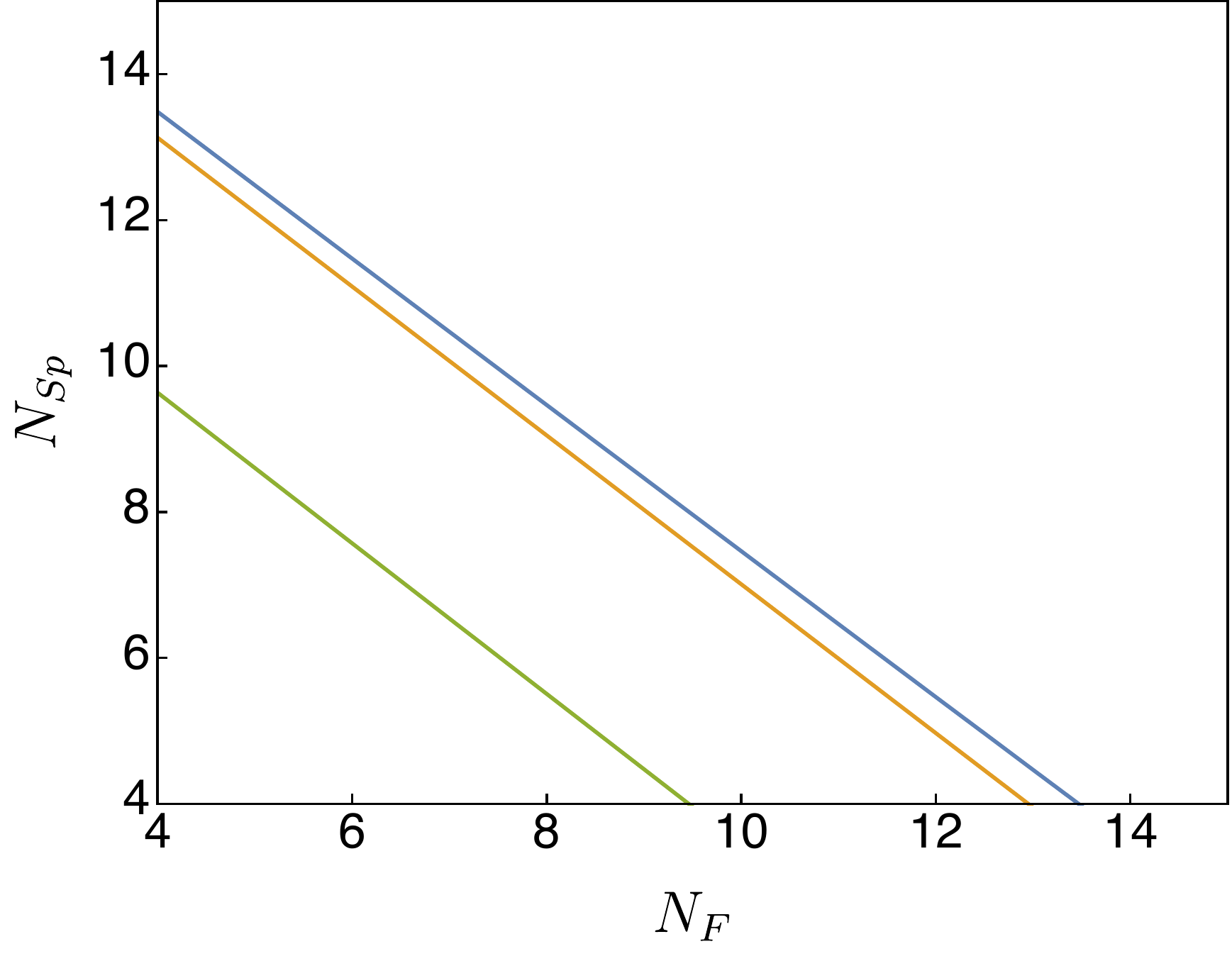}
\includegraphics[width=.47\textwidth]{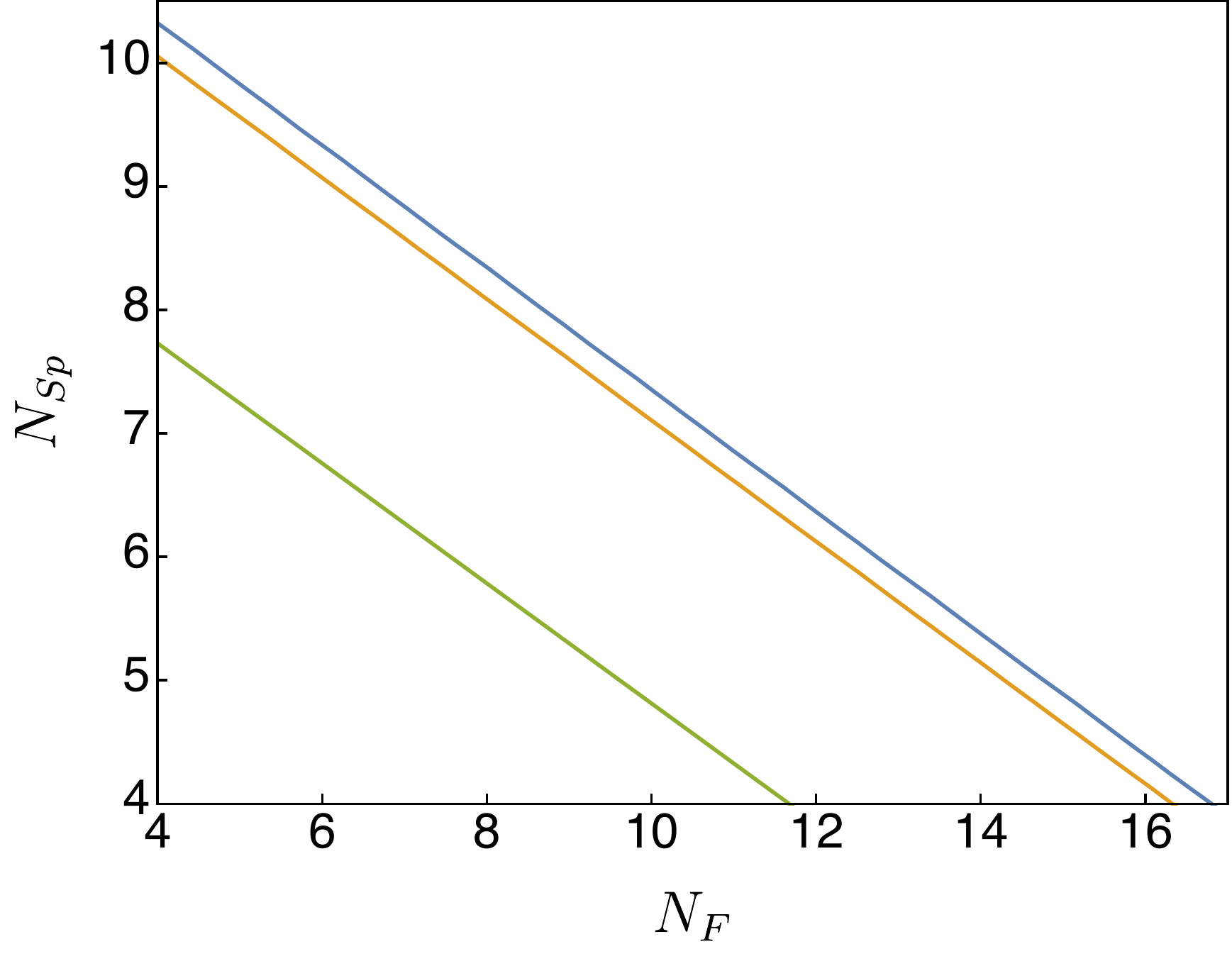}
\includegraphics[width=.47\textwidth]{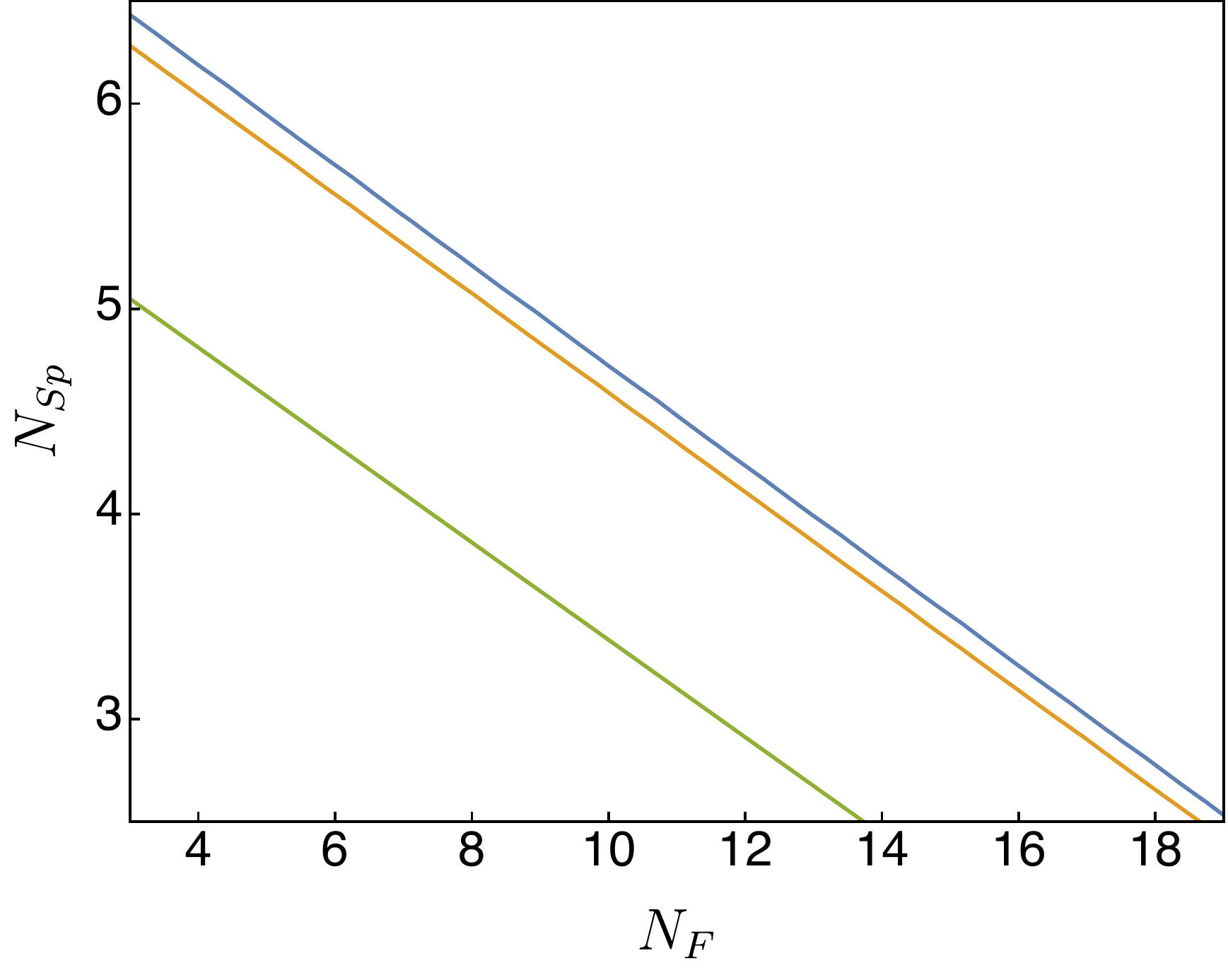}
\includegraphics[width=.47\textwidth]{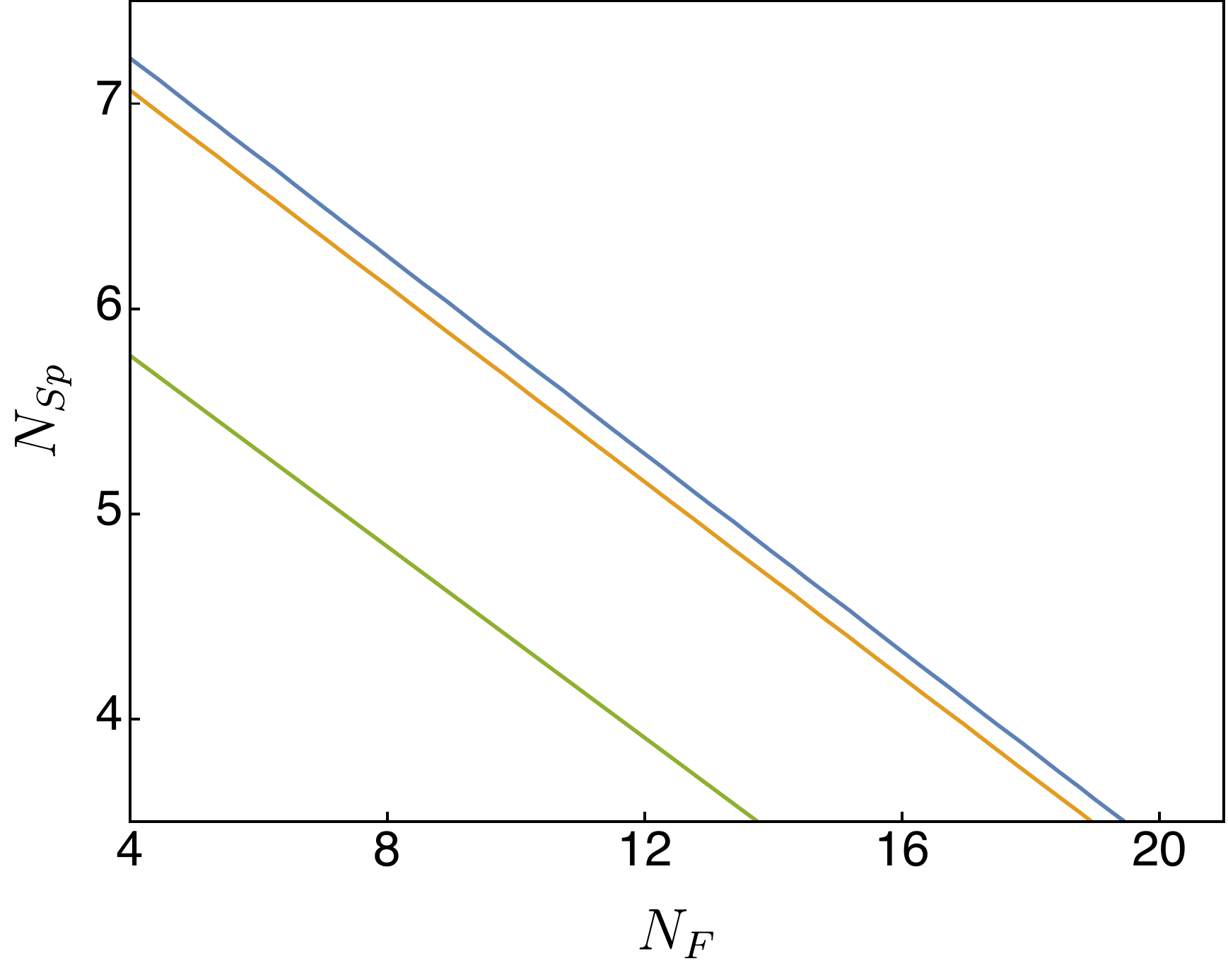}
\caption{%
\label{fig:so7_loops}%
Estimation of the lower edge of the conformal window in $SO(7)$, $SO(9)$, $SO(10)$ and $SO(11)$ gauge theories 
(from left-top to right-bottom) 
containing $N_{\rm F}$ fundamental and $N_{\rm Sp}$ spinorial flavors. 
We use the critical condition $\gamma$CC for the anomalous dimension of the representation $\chi$ at an IR fixed point, $\gcir (2-\gcir)=1$, 
and its scheme-independent series expansions truncated at $\Delta_\chi$ (green), $(\Delta_\chi)^2$ (yellow) and $(\Delta_\chi)^3$ (blue) orders, 
where we identify $R_\chi={\rm \bf F}$ for $SO(7)$ and $R_\chi={\rm \bf Sp}$ for the rest of gauge theories. 
}
\end{center}
\end{figure}

In Figs. \ref{fig:sp4_loops}, \ref{fig:su4_loops} and \ref{fig:so7_loops}, 
we present the critical values $(N_\psi^c,\,N_\chi^c)$, by treating them as continuous variables, 
corresponding to the lower edge of the conformal window 
estimated by applying the scheme-independent critical condition $\gamma$CC 
to two-representation gauge groups discussed in \Sec{CHM}. 
In the figures, green, yellow and blue solid lines denote for the results 
obtained from the scheme-independent series expansions truncated at $\Delta_{\chi}$, $(\Delta_{\chi})^2$ and $(\Delta_{\chi})^3$ orders, 
respectively. 
Note that in $Sp(4)$, $SU(4)$ and $SU(5)$ theories we identify $R_\psi={\rm \bf F}$ and $R_\chi={\rm \bf A2}$, 
in $SO(7)$ theory $R_\psi={\rm \bf Sp}$ and $R_\chi={\rm \bf F}$, 
and in $SO(9)$, $SO(10)$ and $SO(11)$ theories $R_\psi={\rm \bf F}$ and $R_\chi={\rm \bf Sp}$. 

\end{document}